\documentclass[graphics,floatfix, footinbib,tightenlines,nobibnotes, aps, prx, twocolumn]{revtex4-2}
\usepackage{amsmath,amssymb,wasysym}
\usepackage{hyperref}
\hypersetup{
colorlinks = true,
linkcolor = [rgb]{0.70,0.13,0.13},
citecolor = [rgb]{0.13,0.55,0.13},
urlcolor = [rgb]{0.25, 0.41, 0.88}}
\usepackage{color}
\usepackage{xcolor}
\usepackage{braket}

\usepackage[utf8]{inputenc}
\usepackage{bm}
\usepackage{verbatim}
\usepackage{graphicx}
\usepackage{amsmath}
\usepackage{color}
\usepackage{float}
\usepackage{bbm}
\usepackage{amssymb}
\usepackage{slashed}
\usepackage{wasysym,bm,bbm,dsfont,braket}

\begin{document}
\title{Doping a spin-one Mott insulator: possible application to  bilayer nickelate}
\author{Hanbit Oh}
\email{hoh22@jh.edu}
\author{Hui Yang}
\email{huiyang.physics@gmail.com}
\author{Ya-Hui Zhang}
\email{yzhan566@jhu.edu}
\affiliation{William H. Miller III Department of Physics and Astronomy, Johns Hopkins University, Baltimore, Maryland, 21218, USA}

\date{\today}
\begin{abstract}
In this article, we review some recent theoretical developments on potential high-temperature superconductors and unconventional metallic states that can arise from doping a spin-one Mott insulator in the $d^{8}$ valence. These studies are particularly relevant—though not limited—to the recently discovered bilayer nickelate superconductor La$_3$Ni$_2$O$_7$.  We focus on a \textbf{ferromagnetic (FM) Kondo lattice model} with mobile electrons in the $d_{x^2-y^2}$ orbital coupled to the localized spin moments in $d_{z^2}$ orbital through a large Hund's coupling $J_H$. In the large $J_H$ limit, the model reduces to the  \textbf{type II t-J model} with a mixture of \textit{spin-half} singlon states and \textit{spin-one} doublon states.  We summarize DMRG results on the Luther-Emery liquid in one dimensional chain and two-leg ladder. Then we mainly focus on bilayer square lattice and show that a large inter-layer coupling $J_\perp$ of $d_{z^2}$ orbital can induce strong inter-layer pairing of $d_{x^2-y^2}$ orbital. In the strong $J_\perp$ limit, a kinetic-energy driven high $T_c$ superconductivity is demonstrated in an ideal model with only a single hopping term. Furthermore, the model predicts a symmetric pseudogap metal—dubbed ``second Fermi liquid"—in the underdoped regime, yielding a phase diagram analogous to that of hole-doped cuprates. The bilayer Kondo model therefore presents a promising platform for both realizing higher-Tc superconductors and exploring non-Fermi liquid physics. We also comment on the possible limitations of the current models for the bilayer nickelate material and point out some future directions.
\end{abstract}
\maketitle

\section{Introduction}
The problem of doping a spin-1/2 Mott insulator~\cite{lee2006doping} has been extensively studied over the past decades, largely motivated by its relevance to high-temperature superconducting cuprates. Despite intensive theoretical and numerical studies~\cite{lee2006doping,RevModPhys.66.763,RevModPhys.84.1383,PhysRevB.92.195139,PhysRevB.90.165120,Keimer2015,annurev:/content/journals/10.1146/annurev-conmatphys-031620-102024,annurev:/content/journals/10.1146/annurev-conmatphys-090921-033948,PhysRevB.40.506,PhysRevB.62.R9283,doi:10.1126/science.235.4793.1196,PhysRevB.37.3759,PhysRevLett.63.1288,PhysRevLett.58.2691,PhysRevB.38.5142,PhysRevB.35.8865,PhysRevLett.96.047005,Berg_2009,RevModPhys.87.457,PhysRevB.99.235117,PhysRevB.53.251,PhysRevB.62.R14633,PhysRevB.102.115136,PhysRevB.55.R14701,PhysRevB.60.R753,Huang2018,PhysRevB.98.140505,PhysRevB.95.155116,PhysRevResearch.2.033073,PhysRevLett.127.097002,PhysRevLett.127.097003,doi:10.1073/pnas.2109978118,PhysRevX.4.031040,Jiang2021,PhysRevLett.125.157002,PhysRevLett.119.067002,https://doi.org/10.1002/qute.202000126,PhysRevLett.88.117001,PhysRevB.64.100506,PhysRevB.79.220504,PhysRevLett.113.046402,PhysRevLett.122.167001,PhysRevLett.80.1272,grusdt2018meson,chen2018two,zhu2015quasiparticle,zhu2015charge,sun2019localization,zhu2016exact,chen2024phase,wang2015variational}, a well-established theory of the pairing mechanism and the nature of the normal state still remains elusive. It is widely believed that the essential physics of the high Tc cuprate is captured by a single-layer one-orbital Hubbard model on square lattice. However, numerical studies reveal a complicated phase diagram of the model, with superconductivity competing with various symmetry breaking orders (such as the stripe order). The exact nature of the ground state depends on details such as the sign of the next-nearest-neighbor hopping $t'$. For the purpose of both searching for higher critical temperature and understanding the pairing mechanism, it is important to find a different model with a cleaner phase diagram and more robust superconductivity.  We thus turn our attention to an  analogous problem of doping a spin-one Mott insulator, which remains relatively unexplored.  In this article, we will review some recent theoretical works on potential high $T_c$ superconductor and unconventional metallic states from doping a spin-one Mott insulator.

Because an electron only carries spin $1/2$, a spin-one moment must be formed by two electrons together. We will consider a concrete example with the spin-one state in the $3d^{8}$ configuration, with the $d_{x^2-y^2}$ and $d_{z^2}$ orbitals each occupied by one electron. For example, a typical spin-one moment is provided by the Ni$^{2+}$ valence~\cite{buyers1986experimental}.  We then consider the doped state  in  the $3d^{7}$ configuration (for example, the Ni$^{3+}$ valence). We assume a relatively large crystal field energy splitting $\Delta_{\mathrm{cf}}$ between the $d_{x^2-y^2}$ orbital and the $d_{z^2}$ orbital such that the doped hole enters the $d_{x^2-y^2}$ orbital. As a result, the $d_{z^2}$ orbital always has one electron and is in an orbital-selective Mott insulating state, which just provides a spin-half moment. Ultimately, this results in mobile electrons in the $d_{x^2-y^2}$ orbital, which is similar to the hole doped cuprates. However,  new physics arises because the mobile electrons now couple to the local moments from the $d_{z^2}$ orbital through a ferromagnetic Kondo coupling $J_K=-2J_H<0$, where $J_H$ is the Hund's coupling between the two orbitals. Ferromagnetic Kondo lattice model has already been studied and usually hosts a ferromagnetic phase due to the double-exchange mechanism~\cite{PhysRev.118.141}.  In this article, we focus on  the special cases with superconducting ground state, which were not well explored in the early studies. If $J_H$ is large, we will also use a simplified $t-J$ model with a mixture of spin-one state and spin-half state. Such a model was dubbed  type II t-J model~\cite{zhang2020type}.  The major goal of this article is to summarize some existing theoretical results on the FM Kondo lattice model and type II t-J model, with a focus on  the bilayer square lattice.

We start from the one dimensional case, where we dope the familiar Haldane spin-one chain~\cite{PhysRevLett.50.1153}.  The problem with one hole has been studied in Ref.~\cite{penc1995propagating}.  The case of finite density of holes was simulated using density matrix renormalization group (DMRG)~\cite{white1992density,white2023early} in Ref.~\cite{ammon2000spin,Patel2020,PhysRevB.106.045103,PhysRevB.110.064515}. Notably, Ref.~\cite{PhysRevB.106.045103} reported a Luther-Emery liquid phase with pair density wave (PDW) pairing correlations at momentum $Q=\pi$ in both the FM Kondo lattice model and the type II t-J model. A Luther-Emery liquid with uniform pairing can also be found in the two-leg ladder version of the type II t-J model. These two cases have a common feature that the parent spin-one Mott insulator is in a paramagnetic phase with a spin-gap. On the other hand, no superconductivity has been found in square lattice, where the parent insulator is in the antiferromagnetic (AF) ordered phase. Intuitively, magnetism usually competes with the superconductivity and the spin-one magnetism is stronger than the spin-half case. Hence we need another mechanism other than doping to suppress magnetism and promote superconductivity.  Bilayer  version of the FM Kondo model or type II t-J model offers a natural way to suppress magnetism by a stronger inter-layer super-exchange coupling $J_\perp$. Interestingly, a large effective inter-layer spin coupling $J_{\perp;\mathrm{eff}}$ for $d_{x^2-y^2}$ orbital can be mediated mainly by the localized $d_{z^2}$ orbital~\cite{oh2023type,lu2023interlayer}, while the mobile electron in the $d_{x^2-y^2}$ orbital experience a small inter-layer hopping $t_\perp \sim 0$. Such an unusual model with inter-layer spin coupling but vanishing inter-layer hopping turns out to be ideal for realization of a robust superconductor with inter-layer pairing~\cite{oh2023type,lu2023interlayer,yang2024strong}. Similar physics has been proposed in the cold atom system with a bilayer optical lattice~\cite{bohrdt2022strong}, but needs to be out of equilibrium to achieve a finite $J_\perp$ without $t_\perp$.

The models in bilayer square lattice may capture the essential physics in the recently found bilayer nickelate superconductor La$_3$Ni$_2$O$_7$~\cite{sun2023signatures}, which has already attracted lots of experimental~\cite{sun2023signatures,Zhang2024_str,Hou_2023,PhysRevX.14.011040,wang2023observation,ZHANG2024147,zhou2024investigationskeyissuesreproducibility,Wang2024,Wang2024_str,10.1093/nsr/nwaf220,PhysRevLett.133.146002,Dong2024,Chen2024,Chen2024_chemi,Wang2025,Li_2024,li2024distinguishing,zhou2024revealing,Yang2024,Khasanov2025,Wang2025,Ko2025,Zhou2025,Huo2025,Liu2025,10.1093/nsr/nwaf205,10.1093/nsr/nwaf253,bhatt2025resolving,wang2025electronic,sun2025observation,li2025enhanced,Hao2025,fan2025superconducting,shen2025anomalous,wang2025electron} and theoretical studies~\cite{lu2023interlayer,yang2024strong,lange2023pairing,luo2023bilayer,zhang2023electronic,huang2023impurity,Zhang2024,Geisler2024,PhysRevMaterials.8.044801,PhysRevB.109.045151,sakakibara2023possible,tian2024correlation,qin2023high,yang2023minimal,zhan2024cooperation,chen2024non,yang2023possible,gu2023effective,liu2023s,shen2023effective,PhysRevB.109.104508,PhysRevB.109.205156,PhysRevB.109.L201124,oh2023type,zhang2023strong,zhu2025quantum,pan2023effect,PhysRevB.110.024514,PhysRevB.110.L060510,PhysRevB.110.104507,PhysRevB.110.094509,lu2023superconductivity,Luo2024,PhysRevB.109.045127,PhysRevB.109.045154,PhysRevLett.133.096002,Ouyang2024,PhysRevB.108.125105,lange2023pairing,cao2023flat,qu2023bilayer,PhysRevB.108.214522,zhang2023trends,PhysRevB.111.014515,PhysRevB.109.115114,PhysRevB.110.205122,PhysRevB.109.L180502,tian2025spin,liu2025origin,liao2024orbital,PhysRevLett.132.126503,yin2025s,PhysRevB.111.104505,kaneko2025t,ji2025strong,Wang_2025,haque2025dft,shi2025theoretical,gao2025robust,le2025landscape,hu2025electronic,shao2024possible,rm9g-8lm1,ushio2025theoretical,duan2025orbital,qiu2025pairing,cao2025strain,shao2025pairing,PhysRevB.111.L020504,xue2024magnetism}. While several comprehensive reviews on this rapidly developing field  already exist~\cite{cpl_41_7_077402,Si2023}, this article offers a specific perspective using a bilayer version of the FM Kondo lattice model. The resulting framework is applicable not only to the known bilayer nickelate system but can also be extended to similar materials yet to be discovered. 
It is convenient to start from a parent spin-one Mott insulator in Ni$^{2+}$ valence (3$d^8$) and then hole dope it to the Ni$^{(2+x)+}$ valence (3$d^{8-x}$). While the current experimental system is at the fixed doping of $x=0.5$, one can still aim to obtain the full phase diagram for any doping level. Among the existing theoretical studies, there are debates on whether the $d_{x^2-y^2}$ or $d_{z^2}$ orbital dominates the pairing and whether the pairing is intra-layer or inter-layer. Although the type II $t-J$ model can be extended to incorporate the doping of the $d_{z^2}$ orbital~\cite{oh2023type}, we restrict our analysis to the case with $d_{z^2}$ always Mott localized and thus only $d_{x^2-y^2}$ orbital hosts the itinerant electrons and contributes to the pairing. The bilayer type II t-J model and FM Kondo model were simulated by DMRG with a small $L_y$~\cite{yang2024strong, PhysRevB.111.L241102}, which found a strong inter-layer pairing if the inter-layer coupling $J_\perp$ is large.  Interestingly, the pairing survives even if an inter-layer repulsive interaction $V>J_\perp$ is added. This suggests that the naive mean field decoupling of $J_\perp$ does not fully capture the pairing mechanism. Besides, with a finite $V$, the pairing gap increases with doping $x$ at small doping levels, in contrast to the conventional  t-J model where pairing gap decreases with the doping.

To have an analytical understanding of the pairing mechanism without a net attractive interaction, we must go beyond simple mean field treatment. It is quite challenging to deal with a generic values of $J_\perp$ and $V$. However, it is simpler to work in the limit of large $J_\perp$ while keeping the ratio $V/J_\perp$ fixed. In this limit, we should view each rung of the bilayer as a super-site and keep only the following six states: a rung-singlet formed by the spin-half states at each layer, a rung-singlet formed by the spin-one states at each layer, and four singly occupied states with total spin $S=\frac{1}{2}$ on each rung.  We can project the original Hamiltonian into this restricted Hilbert space and get a simplified model dubbed as empty-singlon-doublon (ESD) model~\cite{yang2024strong,oh2025hightemperaturesuperconductivitykineticenergy}.  For the special ratio of $V/J$ such that the net interaction $V_{\mathrm{eff}}=0$, there is only the projected hopping term in the final model.  DMRG simulations demonstrate a strong superconductor with $T_c$ reaching order of the intra-layer hopping in this ideal model~\cite{oh2025hightemperaturesuperconductivitykineticenergy}.  In this model it is also possible to do a generalized slave boson mean field analysis which reproduces the numerical results qualitatively~\cite{yang2024strong}.  Especially, the analytical theory points out two different Fermi liquids in the $x\rightarrow 0$ and $x \rightarrow 1$ region, with a jump of Fermi surface volume per flavor by half of the Brillouin zone (BZ) between them.  The Fermi liquid in the $x\rightarrow 0$ regime has a small hole pocket and may be viewed as a pseudogap metal, but without any symmetry breaking or fractionalization.  The ideal model thus offers a good opportunity to study both unconventional normal state and high Tc superconductor in a cleaner setup.  Although the current experimental system is definitely away from this ideal limit, one may still hope that future experiments can approach this model in a different material and hence realize much higher Tc.

The rest of the paper is organized in the following way. 
In Sec.~\ref{Sec2}, we introduce the ferromagnetic Kondo lattice model and the type-II $t$--$J$ model to describe the doped spin-one Mott insulator.  In Sec.~\ref{Sec3}, we turn to the bilayer system and review the magnetic phase diagram of the parent compound. We then discuss superconductivity and the normal state of the doped system based on the bilayer type-II $t$--$J$ model. In Sec.~\ref{Sec4}, we introduce an ideal model (the ESD model) in the strong-coupling limit. Remarkably, this simple model captures the key physics of the doping-induced superconducting dome and the emergence of a second Fermi liquid in the underdoped normal state. In Sec.~\ref{Sec5}, we review the current experimental and theoretical results on bilayer nickelates, and discuss several open questions that connect idealized theoretical models with realistic materials. Finally, we conclude the article and  highlight potential future directions.

\section{General model: ferromagnetic Kondo model and type II t-J model}
\label{Sec2}
In this section, we review the general model for doping a S=1 Mott insulator, which should work for any lattice and dimension. We start from a two-orbital Hubbard model formed by the two $e_g$ orbitals, $d_{x^2-y^2}$ and $d_{z^2}$ orbitals. We label these two orbitals as $d_{1}$ and $d_{2}$, respectively.

The two-orbital Hubbard model is in the form
\begin{align}
    H=H_K+&\frac{U_1}{2}\sum_{i}n_{1;i}(n_{1;i}-1)+\frac{U_2}{2}\sum_{i}n_{2;i}(n_{2;i}-1)\nonumber\\
+&U^\prime\sum_{i}n_{1;i}n_{2;i}-2J_H\sum_{i}(\vec{S}_{1;i}\cdot\vec{S}_{2;i}+\frac{1}{4}n_{1;i}n_{2;i}),
\label{Eq:two_orbital_Hubbard}
\end{align}
where $n_{a;i}$ is the electron density of $d_a$ orbital at site $i$. $U_{1,2}$ is the intra-orbital interactions and $U^\prime$ is the inter-orbital interaction between the two orbitals. $J_H$ is the Hund's coupling between the two orbitals. One usually expects that $U_1=U_2=U$ and $2J_H=U-U^\prime$. The kinetic energy is  
\begin{align}
H_K &=
     -\sum_{\langle i,j\rangle,\sigma} (t_1 d^\dagger_{1;i;\sigma} d_{1;j;\sigma} + t_{2} d^\dagger_{2;i;\sigma} d_{2;j;\sigma} + h.c. )\nonumber \\
   &+ \sum_{\langle i,j\rangle,\sigma} t_{12;ij} d^\dagger_{1;i;\sigma} d_{2;j;\sigma} + \text{h.c.} 
   + \sum_{i} \Delta_{\mathrm{cf}}\, n_{2;i},
\end{align}
where $t_{1}$ and $t_{2}$ denote the in-plane intra-orbital hoppings for orbitals $d_1$ and $d_2$, respectively. $t_{12;ij}$ represents the in-plane inter-orbital hopping which has a form factor that encodes the distinct symmetries of $d_{x^2-y^2}$ and $d_{z^2}$ orbitals. For instance, if we consider a simple square lattice, $t_{12;ij} = (-1)^{s_{ij}} t_{12}$, with $s_{ij} = 1$ for an $x$-bond and $s_{ij} = -1$ for a $y$-bond. The parameter $\Delta_{\mathrm{cf}}$ denotes the crystal-field splitting between the two orbitals and we mainly consider the case of $\Delta_{\mathrm{cf}}>0$. Throughout this paper, we consider the total filling $n_T = n_{1} + n_{2} = 2 - x$ per site, focusing the range of $0<x<1$.

\begin{figure}
    \centering
    \includegraphics[width=\linewidth]{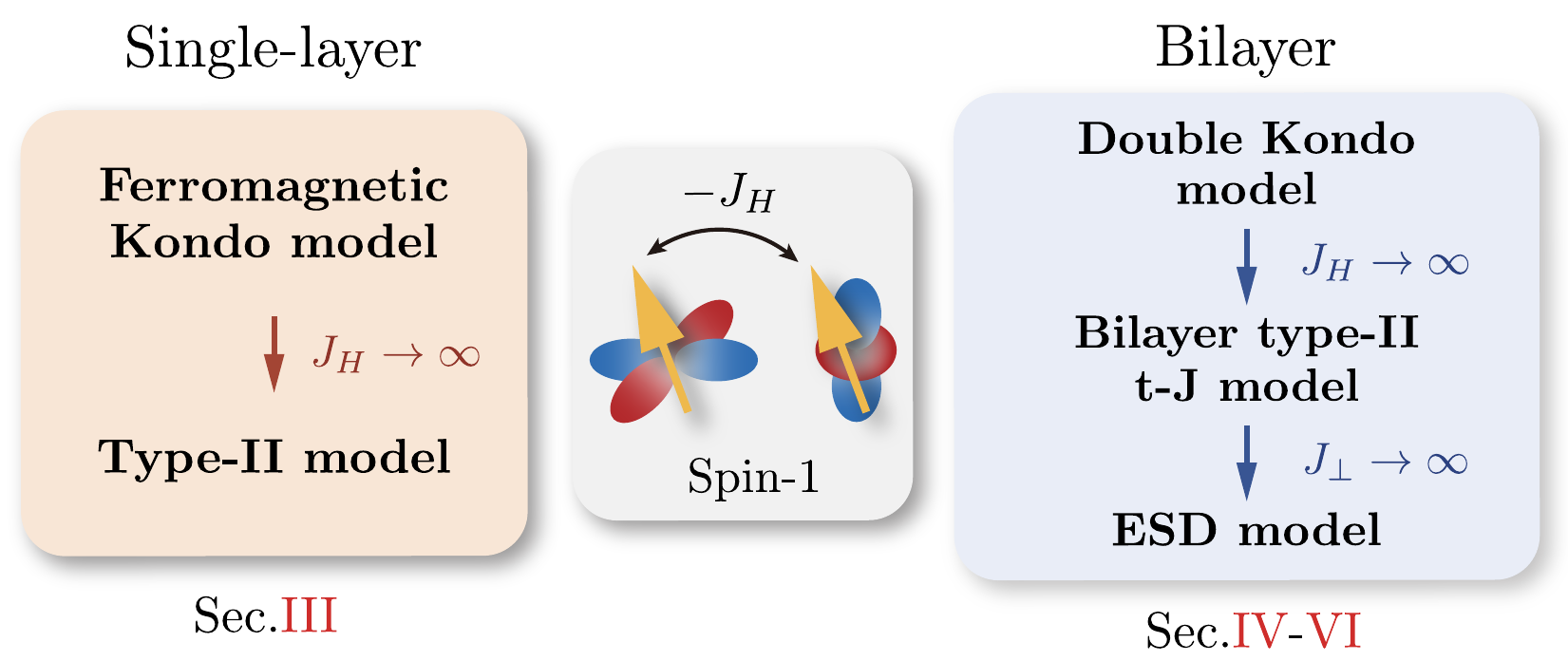}
\caption{\textbf{Doping a spin-one Mott insulator and the relevant theoretical models are central themes of this review.} The spin-one moment arises from two electrons occupying the $d_{x^2-y^2}$ and $d_{z^2}$ orbitals in the $3d^8$ valence configuration. In what follows, we first discuss the single-layer system and then turn to the bilayer structure.
    }
    \label{fig:placeholder}
\end{figure}

\subsection{Ferromagnetic Kondo model}

We start from $x=0$. In this case we expect a Mott insulator with $n_1=n_2=1$ per site when $\Delta_{\mathrm{cf}}<U-U'+J_H$. Then the Hund's coupling $J_H$ enforces the two electrons to form a spin-one moment together.  Then, at a finite doping of $x>0$, holes primarily enter the $d_1$ orbital ($n_1=1-x$), leaving the $d_2$ orbital still at half-filling per spin ($n_2=1$). In this situation, the $d_2$ orbital can be regarded as a localized magnetic moment, and the resultant Hamiltonian becomes a FM Kondo model,
\begin{align}
    H=-&t_0\sum_{\langle i,j\rangle,\sigma}Pc^\dagger_{i;\sigma}c_{j;\sigma}P+h.c.+J_K\sum_{i}\vec{S}_{i}\cdot\vec{s}_{i;c}\nonumber\\
    +&J_c\sum_{\langle i,j\rangle}\vec{s}_{i;c}\cdot\vec{{s}}_{j;c}+J_s\sum_{\langle i,j\rangle}\vec{S}_{i}\cdot\vec{S}_j\nonumber\\
    +&J_{cs}\sum_{\langle i,j\rangle}(\vec{S}_i\cdot\vec{s}_{j;c}+\vec{s}_{i;c}\cdot\vec{S}_j),
    \label{eq:Kondo_model}
\end{align}
where $P$ is a projection operator to remove the double occupancy  for the $d_1$ orbital at each site. $c=d_1$ is an annihilation operator of itinerant electron. The spin operator of the itinerant electron is defined as $\vec{s}_{c;i} = \frac{1}{2} \sum_{\sigma,\sigma'} c^\dagger_{i;\sigma} \vec{\sigma}_{\sigma \sigma'} c_{i;\sigma'}$, where $\vec{\sigma}$ represents the Pauli matrices with $\sigma = \uparrow, \downarrow$ and the spin operator of the localized moment of $d_{2}$ orbital in Eq.(\ref{eq:Kondo_model}) is represented as $\vec{S}_{i}$ at site $i$. The parameters are derived as $t_0=t_{1}$, $J_K=-2J_H$, $J_c=\frac{4t_1^2}{U_1}$, $J_s=\frac{4 t_2^2}{U_2}$ from Eq.(\ref{Eq:two_orbital_Hubbard}). $S$ and $s_{c}$ correspond to the spin operators of localized moment and itinerant electron. $J_{cs}=\frac{2t_{12}^2}{U-U^\prime-\Delta_{\mathrm{cf}}}+\frac{2t_{12}^2}{U+U^\prime+\Delta_{\mathrm{cf}}}$ describes the spin-spin interaction of mobile electron (C-layer) and localized moment (S-layer). Note one difference from the usual Kondo lattice model is that the itinerant electron is itself strongly correlated and described by a $t-J$ model.

\subsection{Type-II t-J model}
The local Hilbert space of the FM Kondo model at each site $i$ is six-dimensional, from the tensor product of three states from the itinerant electron and two states from the localized spin state. In the strong ferromagnetic coupling limit, $J_K=2J_H \to -\infty$, the spin-singlet configuration of $n_1=n_2=1$ can be removed, leaving only five ($5=2+3$) low-energy states grouped as follows~\cite{zhang2020type}:

\begin{itemize}
    \item \textbf{Two singlons} $\ket{\sigma}$: These are $S = 1/2$ states ($\sigma = \uparrow, \downarrow$)  defined as  
    $\ket{\sigma} \equiv \ket{0}_
    C\otimes \ket{\sigma}_S$,  
    where $\ket{0}_c$ indicates the empty state of $d_1$ orbital and $\ket{\sigma}_S$ is the spin state of localized spin in $d_2$ orbital.
    
    \item \textbf{Three doublons}  $\ket{\alpha}$: These are $S=1$ spin-triplet states ($\alpha = 1, 0, -1$) with one electron in both itinerant electron and localized spin, defined as  
    $\ket{1} = \ket{\uparrow}_C \otimes \ket{\uparrow}_S$,  
    $\ket{0} = \frac{1}{\sqrt{2}}(\ket{\uparrow}_C \otimes \ket{\downarrow}_S+ \ket{\downarrow}_C \otimes \ket{\uparrow}_S)$,  
    $\ket{-1} = \ket{\downarrow}_C \otimes \ket{\downarrow}_S$.
\end{itemize}
Projecting the FM Kondo model in Eq.~\ref{eq:Kondo_model} into this $5$-dimensional Hilbert space, we arrive at the type-II t-J model\cite{zhang2020type}:
\begin{align}
    H=-&t_0\sum_{\langle i,j\rangle,\sigma}Pc^\dagger_{i;\sigma}c_{j;\sigma}P+h.c.+J_{ss}\sum_{\langle i,j\rangle}\vec{S}_i\cdot\vec{S}_j\nonumber\\
    +&J_{dd}\sum_{\langle i,j\rangle}\vec{T}_i\cdot\vec{T}_j+J_{sd}\sum_{\langle i,j\rangle}(\vec{T}_i\cdot\vec{S}_j+\vec{S}_i\cdot\vec{T}_j)
    \label{eq:typeII}
\end{align}
Compared to the conventional $t$–$J$ model, the type-II model hosts a mixture of spin-half and spin-one states. Consequently, the model involves two types of spin operators, the spin-$\tfrac{1}{2}$ operator $ \vec{S}_{i} = \frac{1}{2} \sum_{\sigma\sigma'} \vec{\sigma}_{\sigma\sigma'}\, \ket{\sigma}_{i} \bra{\sigma'}_{i} $, and the spin-$1$ operator $ \vec{T}_{i} = \sum_{\alpha,\beta} \vec{T}_{\alpha\beta}\, \ket{\alpha}_{i} \bra{\beta}_{i} $. There are three distinct exchange couplings: $J_{ss}$ (between two singlons), $J_{dd}$ (between two doublons), and $J_{sd}$ (between a singlon and a doublon). They are related to the parameters in Eq.~\ref{eq:Kondo_model} by $J_{ss} = J_s$, $J_{sd} = \frac{1}{2}(J_s + J_{cs})$, and $J_{dd} = \frac{1}{2\sqrt{2}}(J_c + J_s + J_{cs})$.

\subsection{Results in 1D chain and two-leg ladder}

\begin{figure}[ht]
    \centering
\includegraphics[width=\linewidth]{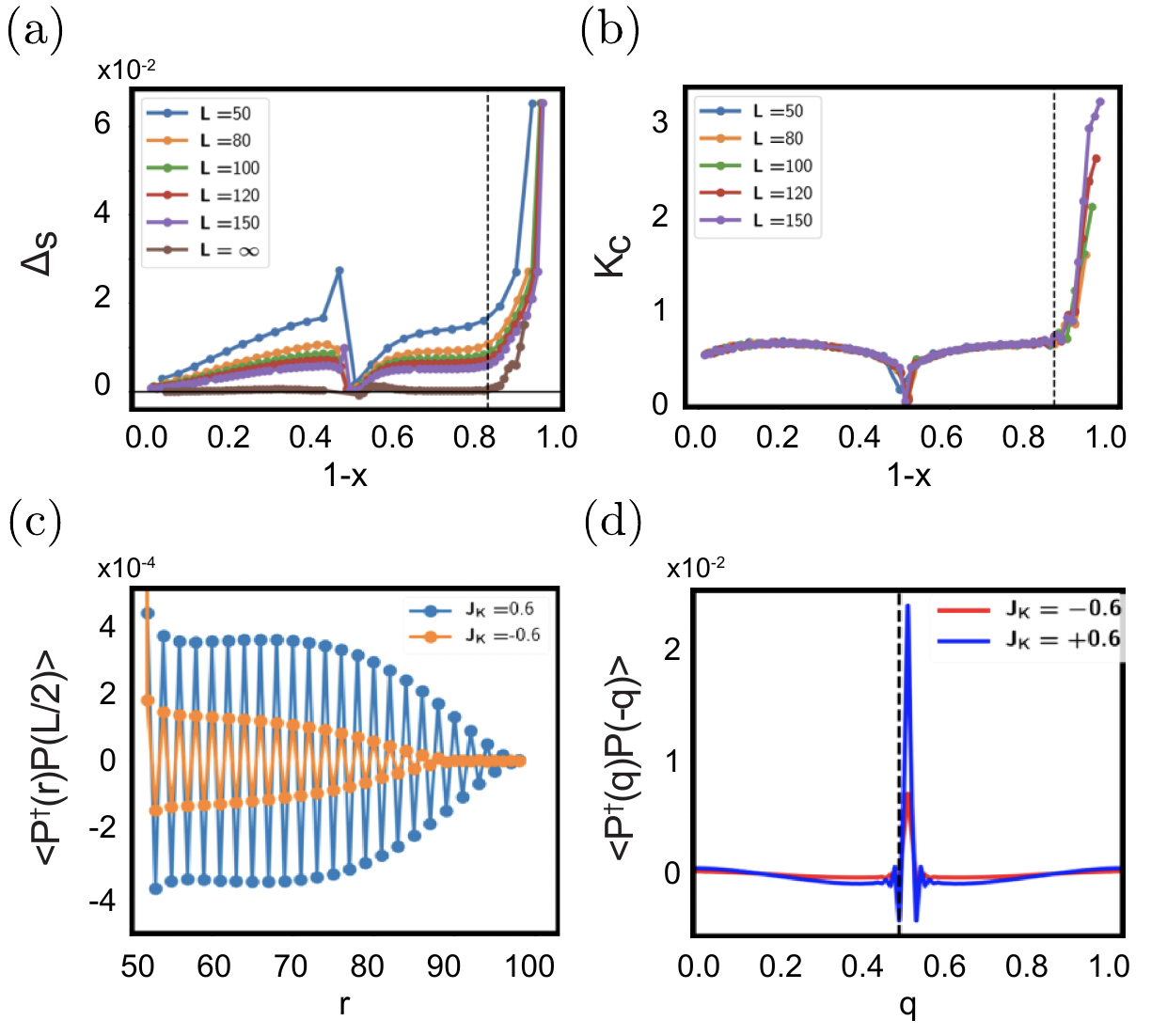}
    \caption{\textbf{Pair-density wave in 1D chain ($L_y=1$) of type-II t-J model.} 
Here, the parameter is set as $t=1, J_{ss}=J_{dd}=0.5, J_{sd}=0.25$.  
(a) Spin gap $\Delta_s$ with $x$. The dashed line is at $x_c=0.15$. $L$ is the system size (b) Luttinger parameter $K_c$ with $x$. The Luttinger parameter $K_c$ becomes large when $x > x_c$, indicating slow decay of pair-pair correlation function. (c,d) The pair-pair correlation function of the Kondo model for $J_K=\pm 0.6, x=0.04$ in real and momentum space, respectively. 
Reprinted from Ref.~\cite{PhysRevB.106.045103}.
  }
\label{fig:type_II_PDW}
\end{figure}

We are now ready to study the FM Kondo model and the type II t-J model in specific lattices. Let us start from the simplest case of doping the one dimensional Haldane chain.  Let us use the type II t-J model in Eq.~\ref{eq:typeII} as an illustration. At zero doping, we just have a Heisenberg model of spin-one moment, whose ground state is known to be in the Haldane phase with a spin gap~\cite{PhysRevLett.50.1153}. Then with slight dope spin-half hole states,  the ground state was shown to be a Luther-Emery liquid with pair-density-wave (PDW) correlations in Ref.~\cite{PhysRevB.106.045103}. As shown in Fig.~\ref{fig:type_II_PDW}, the spin gap decreases with doping $x$, but stays finite below $x_c \approx 0.15$.  Within the spin gapped Luther-Emery phase, there is pair-pair correlation at momentum $Q=\pi$ whose power law exponent is below $1$.  For the region of $x>x_c$, the system has a central charge $c=3$. It smoothly connects to the decoupled phase with a usual Luttinger liquid in the $d_1$ orbital and the gapless spin-half chain from $d_2$ orbital. Interestingly, such a `decoupled' phase survives even in the type II t-J model with strong inter-orbital Hund's coupling $J_H\rightarrow +\infty$. The Luther-Emery liquid phase can be understood within bosonization framework starting from this $c=3$ phase~\cite{PhysRevB.106.045103}. The existence of the Luther-Emery liquid phase was also confirmed in the Kondo model (Eq.~\ref{eq:Kondo_model})~\cite{PhysRevB.106.045103} with both negative and positive $J_K$.  Similar phase was previously also discussed in a Kondo model with positive and large $J_K$~\cite{berg2010pair}.

\begin{figure}[tb]
    \centering
\includegraphics[width=\linewidth]{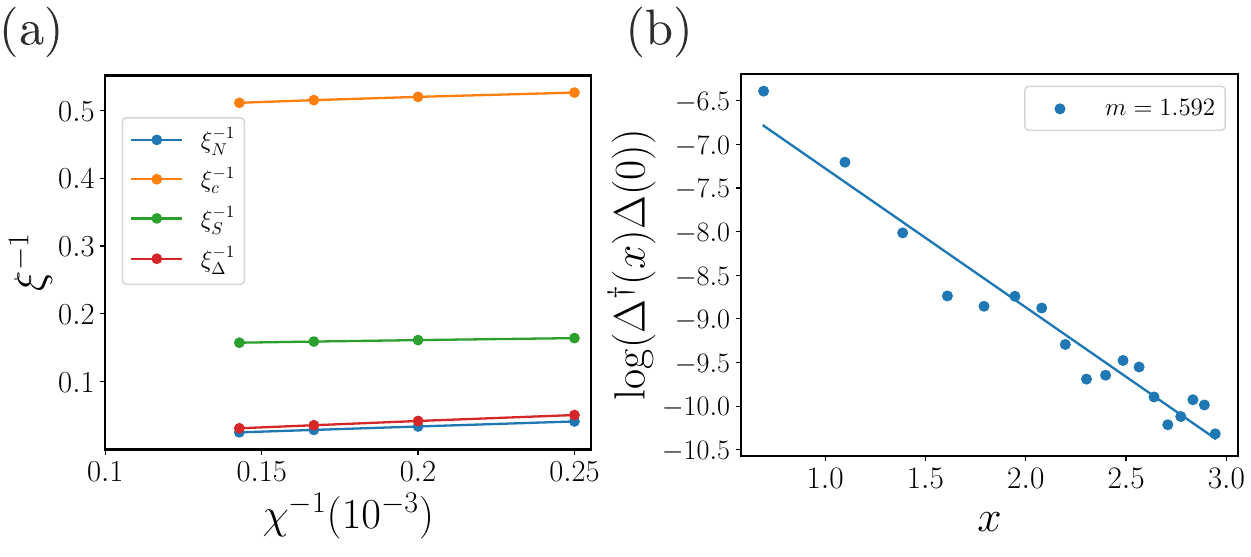}
    \caption{\textbf{Superconductivity in two-leg ladder ($L_y=2$) of type-II t-J model.
} 
    (a) Correlation length of different operators $\xi_N$, $\xi_S$, $\xi_\Delta$, $\xi_C$ at $x=0.1$. The correlation length is calculated from the transfer matrix in the corresponding symmetry sectors. In DMRG simulation, we consider the charge $U(1)$ and spin $S_z$ conservation, the correlation length $\xi_N$, $\xi_{S}$, $\xi_\Delta$, $\xi_C$ correspond to the symmetry sectors with $(N,S_z)=(0,0)$, $(0,1)$, $(1,\frac{1}{2})$, $(2,0)$. (b) Fitted pair-pair correlation function with the power exponent $m=1.592$ in two-leg ladder with $x=0.1$ and $J_{ss}=J_{sd}=0.05,J_{dd}=0.125$.}
\label{fig:superconductivity_two_leg}
\end{figure}

Then we move to  two-leg ladders ($L_y=2$), the type-II t-J model remains in a Luther-Emery liquid phase at small doping $x=0.1$, as shown in Fig.~\ref{fig:superconductivity_two_leg}.  However, now the pair correlation is at zero momentum.  The result is similar to the conventional t-J model and may be understood from doping a parent `rung singlet' of spin-one moments.  In contrast, no superconductor phase was found in DMRG of the type II t-J model in four-leg ladder.  In the 2D limit, the parent Mott insulator is known to be in a antiferromagnetic ordered phase. Given that the magnetism of spin-one moments is generically stronger than that of spin-half moments, it is not clear whether purely doping is enough to suppress the magnetic order and favor a superconductor. To realize a robust superconductor, it is desirable to start from a paramagnetic spin-one Mott insulator. This is quite natural in 1D chain and two-leg ladder, but is challenging in two dimensional square lattice.

\section{Robust superconductivity in bilayer model}
\label{Sec3}

\begin{figure}[tb]
    \centering
\includegraphics[width=0.8\linewidth]{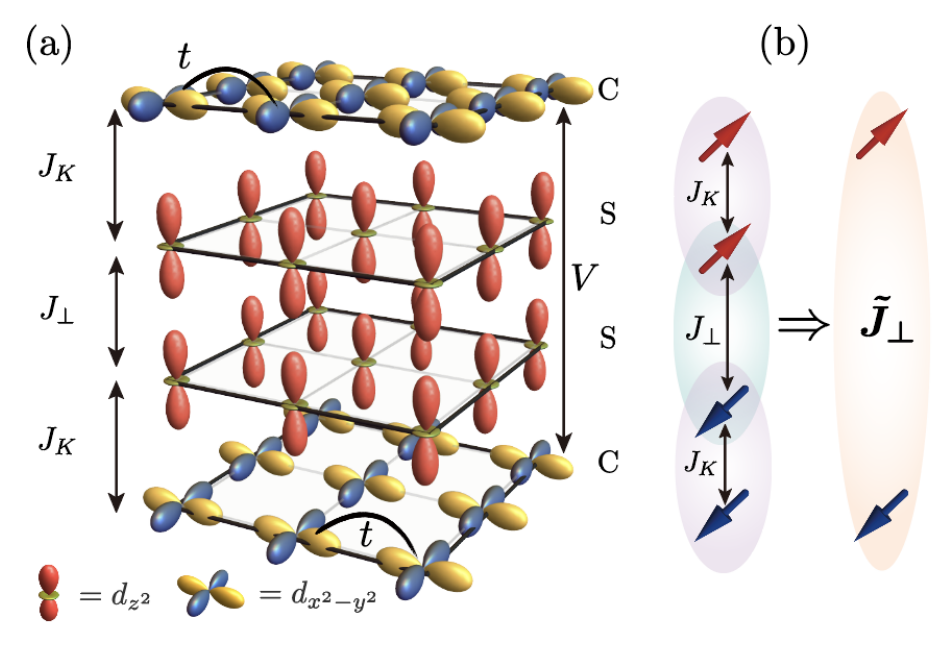}
\caption{\textbf{Illustration of the double-Kondo model.} 
(a) The model consists of two copies of the FM Kondo model. Between the two layers, there is an interlayer exchange coupling $J_{\perp}$ of the S-layer, together with a repulsive Coulomb interaction $V$ within the C-layer. In the context of bilayer nickelates, the $d_{x^2-y^2}$ orbital can be viewed as a mobile conduction electron layer (C-layer), while the $d_{z^2}$ orbital corresponds to a localized moment layer (S-layer). Reprinted from Ref.~\cite{PhysRevB.111.L241102}.
(b) Strong FM coupling $J_K<0$ transmits $J_\perp$ of the $d_{z^2}$ orbital to the $d_{x^2-y^2}$ orbital, $\tilde{J}_\perp$.
    }
    \label{fig:double_kondo_orbital}
\end{figure}

As discussed in the last section, pairing correlation can emerge from doping a spin-one rung singlet state in the two-leg ladder configuration. Intuitively a spin gapped parent state is important to favor the superconductivity. Similar paramagnetic phase is also possible in the bilayer square lattice with a strong inter-layer spin coupling $J_\perp>0$.  We now consider a  bilayer version of the FM Kondo model defined in Eq.~\ref{eq:Kondo_model}, referred to as the \textbf{double-Kondo model} (see Fig.~\ref{fig:double_kondo_orbital}). We label the top and bottom layers as $l = t, b$.  Each layer consists of the two $e_g$ orbitals, which are shown as two separate layers in Fig.~\ref{fig:double_kondo_orbital} just for illustrative purpose.   Again the $d_{x^2-y^2}$ orbital ($d_1$) is described by a $t-J$ model and the $d_{z^2}$ orbital ($d_2$) just provides localized spin moments.   Due to the large inter-layer hybridization of the $d_{z^2}$ orbital, there is a term $J_\perp\sum_i \vec{S}_{i;t}\cdot \vec{S}_{i;b}$  which couples the spin moments of the two layers.  We also include an inter-layer repulsion term $V \sum_i n_{i;t} n_{i;b}$ term. Here $i$ labels the square lattice site.

Within each layer there is also a FM Kondo ocupling $J_K=-2J_H$. When $J_H$ is large,  the low-energy effective theory reduces to the \textbf{bilayer type-II $t$–$J$ model}~\cite{oh2023type}. 
Similar to the singler layer case in Eq.~\ref{eq:typeII}, the model has three types of spin couplings: $J_{ss}$, $J_{sd}$, and $J_{dd}$, but now for both intra-layer and inter-layer bonds.

In the following we summarize some results for the bilayer FM Kondo model and type II  t-J model. 

\subsection{Phase diagram of the $d^8$ parent state}
\label{Sec3A}
\begin{figure}
    \centering
    \includegraphics[width=\linewidth]{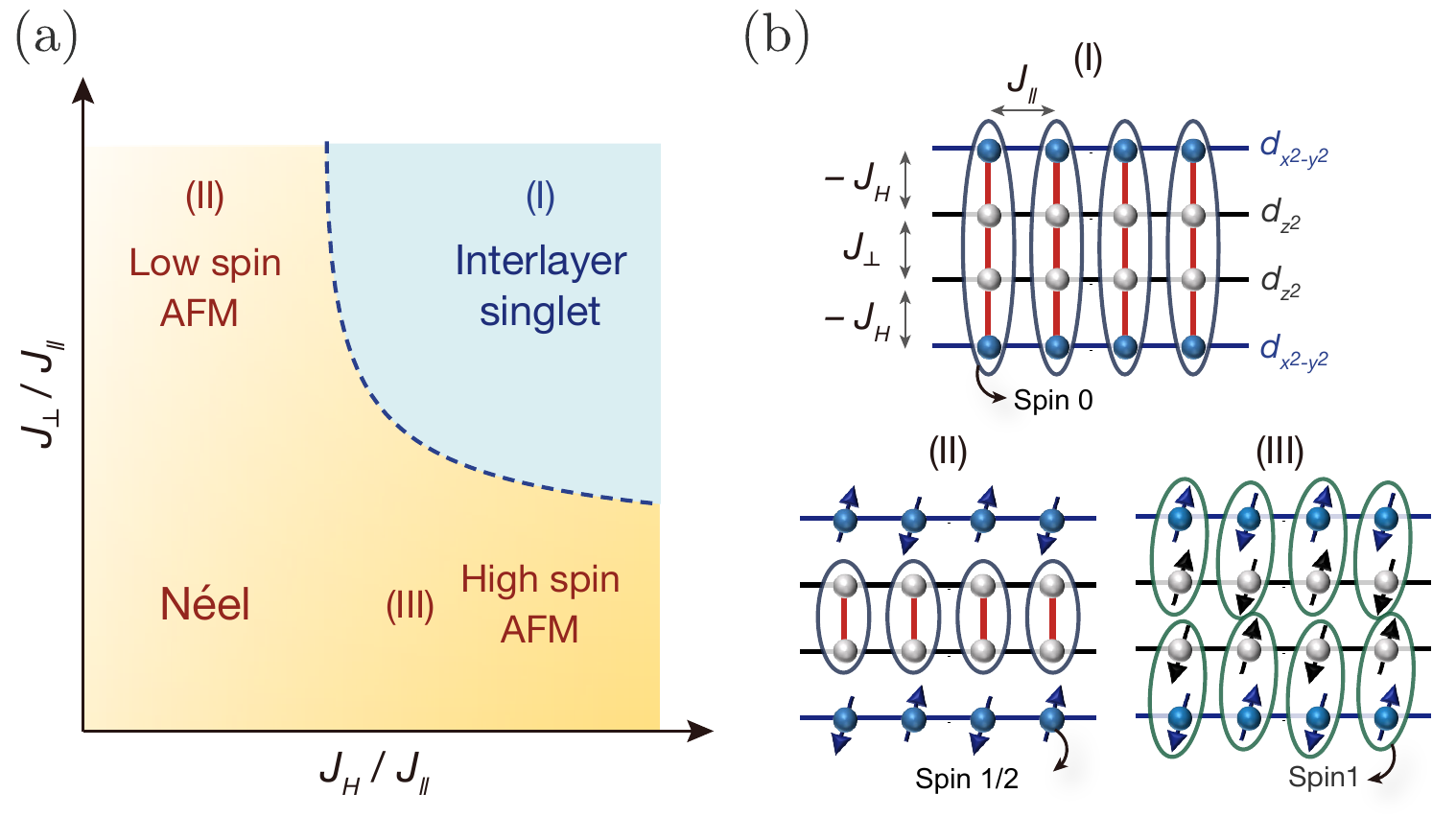}
    \caption{\textbf{Schematic phase diagram and magnetic phases of $J-J_\perp-J_H$ model for parent compounds.} (a) The phase diagram exhibits a second-order phase transition between a N\'eel-ordered phase at small $J_\perp$ and a interlayer singlet phase at large $J_\perp$. 
(b) Illustration of (I) interlayer singlet, (II) low-spin AFM, and (III) high-spin AFM. 
Black ellipses with red lines in (I) and (II) indicate interlayer singlet bonds. Reprinted from Ref.~\cite{oh2025highspinlowspin}.
 }
  \label{fig:d8_phase_diagram}
\end{figure}

For undoped case ($x=0$), the double-Kondo model reduces to a bilayer two-orbital spin-1/2 Heisenberg model~\cite{oh2025highspinlowspin} with three energy scales $(J, J_\perp, J_H)$. 
Thus, even in the undoped case, the interplay between Hund’s coupling and interlayer exchange gives rise to various phases. In the limit $J_H \to 0$, the orbitals decouple. For $J_\perp \gg J_\parallel$, the spins of $d_{2}$ orbital form interlayer singlets, while the spins of $d_1$ orbital develop N\'eel order—defining a \textit{low-spin phase} governed by spin-1/2 moments. In the opposite limit, $J_H \to \infty$ and $J_\perp \to 0$, the two spins align into local triplets, forming effective spin-1 moments—corresponding to the \textit{high-spin phase}. Finally, when both $J_H, J_\perp \gg J_\parallel$, the spin-1 moments form a \textit{interlayer-singlet}, resulting in a gapped, nonmagnetic state. The Ref.~\cite{oh2025highspinlowspin} employed various analytical approaches and characterize the quantum transition between the rung-singlet at large $J_\perp$ and a N\'eel-ordered phase at small $J_\perp$ in this $J$–$J_\perp$–$J_H$ model. In Fig.~\ref{fig:d8_phase_diagram}, the second-order quantum phase transition between the rung-singlet at large $J_\perp$ and a N\'eel-ordered phase at small $J_\perp$ is illustrated. Moreover, in N\'eel-ordered phase, there is the crossover between low- and high-spin states.

The results implies, as long as the Hund coupling and the interlayer coupling are sufficiently large, the rung-singlet state is favored with a spin gap. In the remainder of this paper, we mainly focus on superconductivity arising from doping into the interlayer singlet state. This is in sharp contrast to the single-layer case, where the spin-$1$ N\'eel state is stabilized, which is generally detrimental to superconductivity. In the bilayer system, by suppressing such magnetic competition, superconductivity can be enhanced. In more realistic situations, however, the intralayer exchange $J$ can compete with either $J_\perp$ or $J_H$. Consequently, antiferromagnetic order may compete with superconductivity, a possibility that will be briefly discussed in Sec.~\ref{Sec5}.

\subsection{Inter-layer pairing at finite $x$}

\begin{figure}
    \centering
\includegraphics[width=\linewidth]{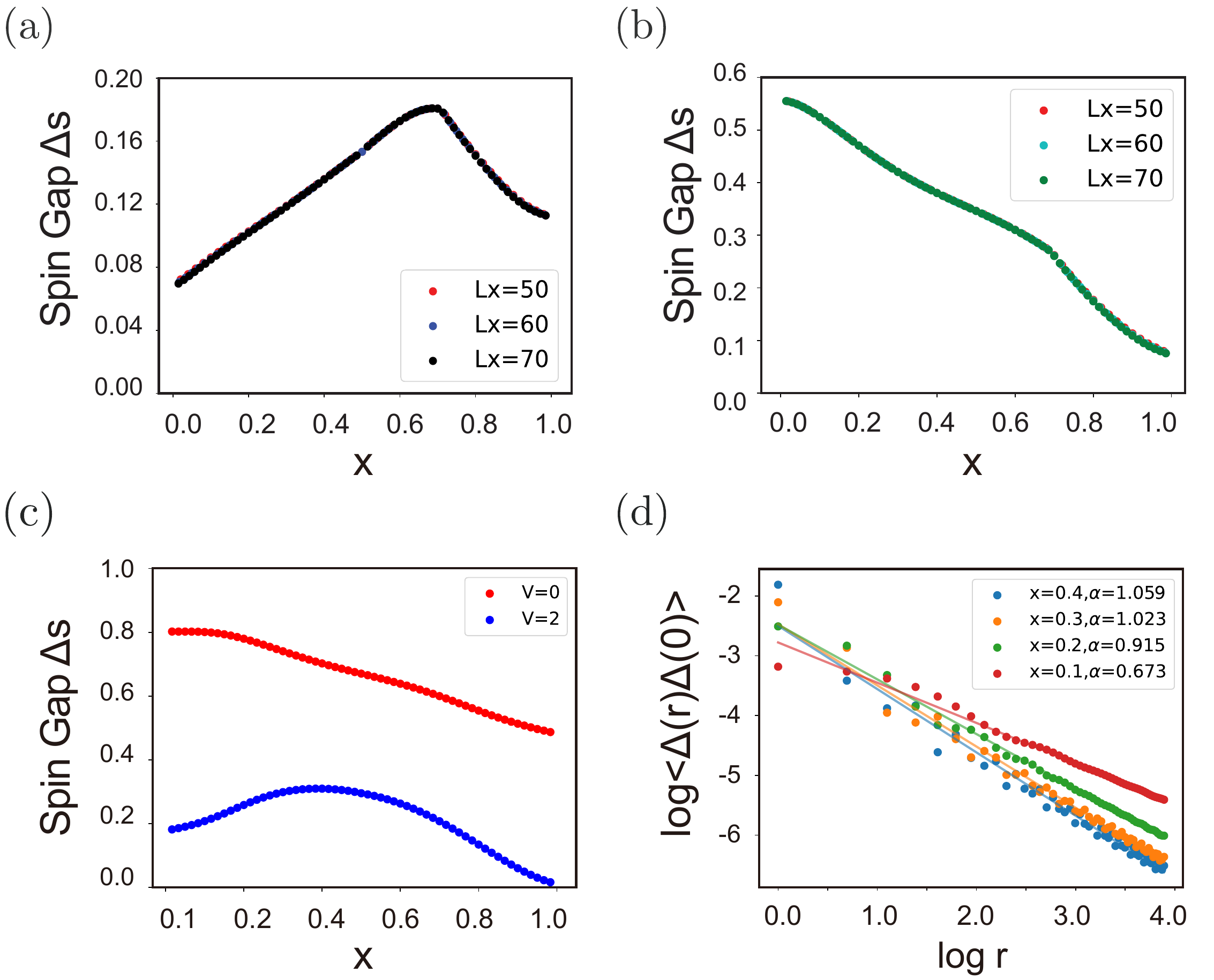}
   \caption{
  \textbf{Evidence of Luther-Emery liquid phase from DMRG simulation of bilayer type-II t-J model for $L_y=1,L_z=2$}.
    (a) The doping dependence of the
    spin gap of the type-II t-J model at $V=0$. 
    We used $J^{ss}_{\perp}=1$, $J_\parallel^{ss}=0.5$ and the relation $J^{ss}=2J^{sd}=4J^{dd}$ holds for both $J_\parallel$ and $J_\perp$. There is a dome structure for the spin gap, and thus the pairing strength. 
    (b) In contrast, for the one-orbital $t-J-J_\perp$ model with $J_\perp=1$ and $J_\parallel=0.5$, the spin gap monotonically decreases with $x$. Here in both (a) and (b) we used bond dimension $\chi=2000$, and the results converge for system size $L_x=50,60,70$. (c) Spin gap in the type-II t-J model with $J_{\perp}^{ss}=5$ and $J_\parallel=0$ for $V=0$ and $V=2$.  One can see that the pairing dome disappears at $V=0$ but comes back at $V=2$.  (d) The pairing correlation functions show the power-law decaying behaviour, $\langle\Delta^\dag(r)\Delta(0)\rangle\sim |r|^{-\alpha}$, where the power law exponent $\alpha$ is smaller than one from iDMRG calculation of the type II t-J model at   $J_{\perp}^{ss}=2$ and $J_\parallel=0$ for the dopings $x=0.1,0.2,0.3,0.4$. The results are consistent with a Luther-Emery liquid with power-law superconductivity pairing correlations.
  Reprinted from Ref.~\cite{yang2024strong}.
    }
    \label{fig:pd_bilayer_typeII}
\end{figure}

Given that the parent state is in a rung singlet phase in the large $J_H, J_\perp$ regime, superconductivity with inter-layer pairing naturally emerges. To confirm this intuition, Ref.~\cite{yang2024strong} analyzed the bilayer type-II $t$–$J$ model, employing both analytical and numerical techniques.  DMRG simulation with $L_y=1, L_z=2$ clearly shows a strong inter-layer pairing. Surprisingly, the pairing gap has a dome structure with the doping $x$, as shown in Fig.~\ref{fig:pd_bilayer_typeII}(a)(c).  This is in contrast to the cuprate phase diagram where pairing gap decreases with hole doping.  Moreover, the pairing survives with a large $V$ such that the net interaction is repulsive. Therefore, a simple mean field decoupling of the $J_\perp$ term may not fully capture the pairing mechanism.

Interlayer pairing has also been found in the simplified bilayer one-orbital $t-J$ model, which consists of only the $d_{x^2-y^2}$ orbital and a phenomenological interlayer coupling, $J_{\perp; \mathrm{eff}}$~\cite{lu2023interlayer,oh2023type,qu2023bilayer,lu2023superconductivity,lange2023pairing,PhysRevB.109.045127}. While such a model may be relevant for certain physical systems like bilayer optical lattices~\cite{bohrdt2022strong}, its applicability to the bilayer two-orbital Hubbard model or bilayer FM Kondo model is limited to a specific parameter regime. The one-orbital description is rigorously justified only in the $J_H \ll J_\perp$ limit, where the $d_{z^2}$ orbital can be integrated out to produce a small effective spin coupling $J_{\perp;\mathrm{eff}}\sim \frac{J_H^2}{J_\perp}$ for the $d_{x^2-y^2}$ orbital. In the large $J_H$ regime central to this article, this approximation breaks down because the spin of the $d_{z^2}$ orbital becomes an active low-energy degree of freedom. Although this simple model correctly predicts the pairing symmetry, its quantitative results diverge significantly from the more complete bilayer type II $t-J$ model (see Fig.~\ref{fig:pd_bilayer_typeII}(a,b)). Indeed, a naive one-orbital model with an assumed $J_{\perp; \mathrm{eff}}=J_\perp$ significantly overestimates the pairing strength~\cite{oh2025hightemperaturesuperconductivitykineticenergy}, rendering its use in this context questionable.

\section{Ideal model in the large $J_\perp$ limit: kinetic superconductor and second Fermi liquid}
\label{Sec4}

In the last section, we provided numerical results suggesting the possibility of superconductivity in the bilayer type II t-J model. However, due to the computational complexity, the calculation is restricted to small $L_y$ and it is not clear whether the same conclusion holds in the two dimensional (2D) limit. In addition, the DMRG results suggest that pairing survives when the net interaction $V_{\mathrm{eff}}$ is even positive, which does not have a satisfactory analytical understanding.  In the simple mean field treatment~\cite{oh2023type,lu2023interlayer}, one can simply do a mean field decoupling of an effective $J_\perp$ coupling between the $d_{x^2-y^2}$ orbital. However, when a larger $V$ is added, pairing is suppressed if we also include the mean field decoupling of the repulsive $V$ term.  From numerical results~\cite{yang2024strong} the pairing strength decreases with $V$, so the repulsion $V$ term definitely has strong influence on the ground state, but its effect seems to be beyond simple mean field theory analysis. 

To perform numerical simulation in much larger $L_y$ and to have a better analytical description, we will consider a simpler model in the large $J_\perp$ limit of the bilayer Kondo model or bilayer type II t-J model, with the ratio $V/J_\perp$ fixed.  For this ideal model, strong pairing is demonstrated in DMRG with $L_y$ as large as $8$. Meanwhile, at least the qualitative trend of the numerical results can be reproduced in a generalized slave boson mean field theory. 

\subsection{The ESD  model in the strong coupling limit}

\begin{figure}
    \centering
\includegraphics[width=\linewidth]{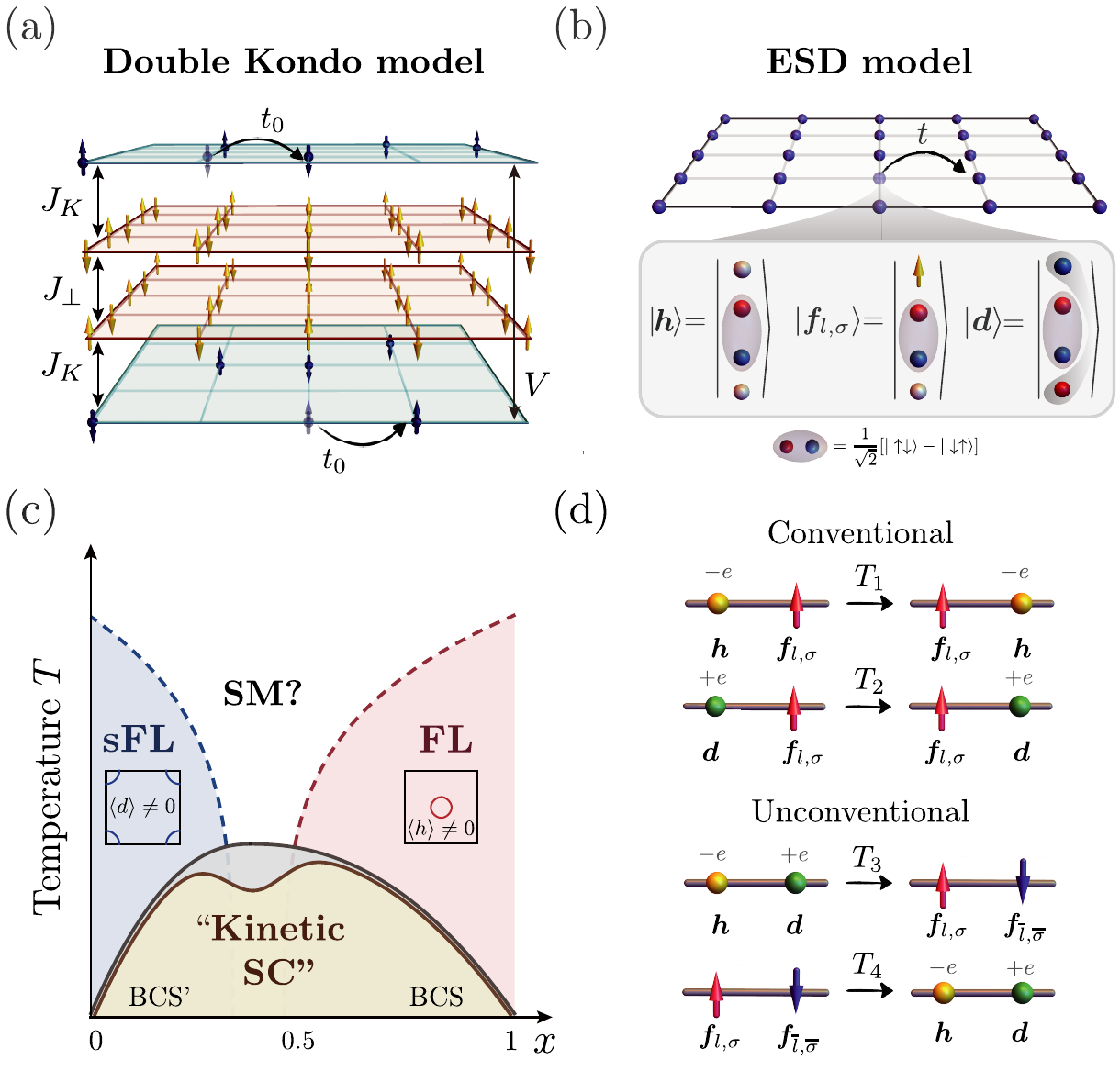}
    \caption{\textbf{Illustration of ESD model and its phase diagram.}
(a,b) In the strong-coupling limit, the double Kondo model reduces to a simpler model, ESD model. The ESD model retains only six states on each rung, which combines the two layers: one empty (E) state $|h\rangle$, four singlon (S) states $|f_{l\sigma}\rangle$, and one doublon (D) state $|d\rangle$. 
(c) Schematic phase diagram of the ESD model.
The solid black and brown lines indicate the pairing and coherence temperature scales, respectively. In our ESD model, a superconducting pairing dome extends from two distinct Fermi liquids: the conventional Fermi liquid (FL) near $x = 1$ and a second Fermi liquid (sFL) near $x = 0$.  (d) Illustrations of four distinct hopping processes between two adjacent sites.
$T_1$, $T_2$ are conventional hoping, while $T_3$, $T_4$ are unconventional one.
 Reprinted from Ref.~\cite{oh2025hightemperaturesuperconductivitykineticenergy}.
}
    \label{fig:esd_pd}
\end{figure}

In the strong coupling limit ($|J_K|, J_\perp,V \gg  \textcolor{black}{t_0}, J_s,J_c$), the double Kondo model reduces to a simpler one. In this limit, we view  each rung (labeled by $i$) as a supersite and the low energy states can be obtained from exact diagonalization of  the $J_K$ and $J_\perp$ terms. We keep only six states  which are classified by the total number of electrons ($n_T$) and the total spin ($S_T$):
\begin{itemize}
    \item \textbf{One empty state $|h\rangle$}: $n_T = 0$, $S_T = 0$.
    \item \textbf{Four singlet states $|l, \sigma\rangle$}: $n_T = 1$, $S_T = 1/2$.
    \item \textbf{One doublon state $|d\rangle$}: $n_T = 2$, $S_T = 0$.
\end{itemize} 
We refer to the model projected onto these six states ($6 = 1 + 4 + 1$) as the \textbf{ESD model} (see Fig.~\ref{fig:esd_pd}(b))~\cite{yang2024strong}. The acronym ESD stands for Empty (E), Singlon (S), and Doublon (D). The doublon state corresponds to the interlayer rung-singlet state discussed in Sec.~\ref{Sec3}~A. The internal structures of these states are complex and  vary with $J_K / J_\perp$. However, within the ESD model, these details are not essential. At leading order, we only project the hopping term and the on-site energy shift to the restricted Hilbert space, the final model is in the form

\begin{align}
\hat{H}_{\mathrm{ESD}} = -t \sum_{l,\textcolor{black}{\sigma}\langle i,j \rangle} P c^\dagger_{i;l;\sigma} c_{j;l;\sigma} P + \textcolor{black}{V_{\mathrm{eff}}} \sum_{i} (n_{h;i} + n_{d;i}),
\label{eq:Ham_esd}
\end{align}
with the density operators $n_{h;i} = |h\rangle_i \langle h|_i$ and $n_{d;i} = |d\rangle_i \langle d|_i$. Here, $\textcolor{black}{V_{\mathrm{eff}}}$ characterizes the net interaction, with $\textcolor{black}{V_{\mathrm{eff}}} < 0$ corresponding to an attractive interaction and $\textcolor{black}{V_{\mathrm{eff}}} > 0$ corresponding to a repulsive interaction. We emphasize that the hopping $t$ is rescaled compared to $t_0$ in the double Kondo model, depending on $J_K/J_\perp$. \textcolor{black}{ We assume mirror reflection symmetry that exchanges the two layer remains unbroken, and therefore the two layers have equals densities.}
Accordingly, the creation (annihilation) operator is given by
\begin{align}
c_{i,l,\uparrow} &= |h\rangle_i \langle l,\uparrow|_i + r |\bar{l},\downarrow\rangle_i \langle d|_i, \label{eq:electron1} \\
c_{i,l,\downarrow} &= |h\rangle_i \langle l,\downarrow|_i - r |\bar{l},\uparrow\rangle_i \langle d|_i. \label{eq:electron2}
\end{align}
where $r$ is a parameter that depends on $J_K/J_\perp$ in the original double Kondo model.

Moreover, we have the constraint $n_{d;i} + n_{f;i} + n_{h;i} = 1$ at each site $i$. On average, we have $\frac{1}{2} n_f + n_d = x$, with the hole doping per site defined as $n_T = 2(1 - x)$. The factor of 2 arises from the two layers, and the density per layer is $1 - x$. From $n_T = n_f + 2n_d$, one can derive $n_T = 1 + n_d - n_h$. Thus, the $d$ and $h$ states can be interpreted as carrying charges $+1$ and $-1$, respectively, while the singlon state is neutral and represents a spin moment. In contrast to the conventional $t$-$J$ model, which has carriers with either $+1$ or $-1$ charge, here we have both. In Fig.~\ref{fig:esd_pd}(d), we illustrate the hopping term. The hopping can be divided into two categories: (1) exchange of a singlon state ($f$) and a nearby $h$ or $d$ state; (2) annihilation of a pair of $d$ and $h$ states to create a pair of $f$ states, and the reverse process. Category (1) also exists in the conventional $t$-$J$ model, but category (2) is unique to our ESD model. As we will see later, this leads to interesting new physics not present in the familiar $t$-$J$ model.

The model has three parameters: $t$, $r$, and $\textcolor{black}{V_{\mathrm{eff}}}$. The parameters $t$ and $r$ are determined by the internal structure of the six states, while $\textcolor{black}{V_{\mathrm{eff}}}$ represents the net interaction arising from both the $J_\perp$ and $V$ terms. Specifically, $\textcolor{black}{V_{\mathrm{eff}}} = \epsilon_0 + V/2$, where
 \begin{eqnarray}
  \epsilon_{0}&=&
   \frac{J_\perp}{4} \left[-1 - 
   \sqrt{1 -2\frac{J_K}{J_\perp}+ 4\left(\frac{J_K}{J_\perp}\right)^2}\right. \nonumber
   \\
   &&\quad 
   \left. +2 \sqrt{1 -\frac{J_K}{J_\perp}+\left( \frac{J_K}{J_\perp}\right)^2}\right].
\end{eqnarray}

For simplicity, we set $t = 1$ throughout this study. We primarily focus on the case where $\textcolor{black}{V_{\mathrm{eff}}} = 0$, implying that the $V$ term exactly cancels the attraction from the $J_\perp$ term. In this case, our ESD model contains only one term: nearest-neighbor hopping.
We also note that, because we consider  $\textcolor{black}{V_{\mathrm{eff}}}\geq 0$, the model is fermionic, and the singlon states play a crucial role. Therefore, it does not reduce to a simple hard-core boson theory justified only in the limit $\textcolor{black}{V_{\mathrm{eff}}} \to -\infty$. \textcolor{black}{The $T_c$ analysis in the hard-core bosonic theory corresponding to $\textcolor{black}{V_{\mathrm{eff}}} \rightarrow -\infty$ in our model is well studied in the previous literature \cite{Schlömer2024,Carrasquilla_2013}.}    From the bilayer type-II $t$–$J$ model, the corresponding parameter in the ESD model is  at $r=\sqrt{2}/\sqrt{3}$. Notably, at $r=1$ the model has an additional particle–hole symmetry. In what follows, we will primarily focus on this symmetric point, $r=1$, but the phase diagram of $r=\sqrt{2}/\sqrt{3}$ is qualitatively similar.

\subsection{Kinetic superconductivity}
In this section, we will discuss the kinetic superconductivity arising from the ESD model.
Based on the conventional wisdom, superconductivity is usually assumed to arise from attractive interaction. 
However, surprisingly, Ref.~\cite{oh2025hightemperaturesuperconductivitykineticenergy} shows that strong pairing is possible solely from kinetic energy even without a net attraction at $V_{\mathrm{eff}}=0$ of the ESD model defined above. 
This approach is akin to the $t$-$J$ model with $J=0$, where kinetic magnetism~\cite{PhysRev.147.392} has been previously studied.

To understand how pairing still arises in this non-attractive interaction setting, it is crucial to separate the two roles of the $J_\perp$ term of the double-kondo model:
(I) $J_\perp$  lowers the average energy of the doublon and holon states relative to the singlon by $\epsilon_0 \propto -J_\perp<0$. This gives an attractive interaction.  With a $V$ term, the net  interaction is $V_{\mathrm{eff}} = V + 2\epsilon_0$ so this effect can be offset by $V$.
(II) $J_\perp$ introduces an energy splitting between the $S = 0$ and $S = 1$ states of the holon and doublon. For example, for the doublon states, the energy difference is $\Delta_d = \frac{1}{4}J_\perp$ in the large negative $J_K / J_\perp$ limit. This splitting penalizes spin-triplet configurations and frustrates single-particle hopping.
Conventional mean-field decoupling relies on the effect (I), and thus predicts superconductivity only for $V_{\mathrm{eff}} < 0$. 
However, Ref.~\cite{oh2025hightemperaturesuperconductivitykineticenergy} shows superconductivity also in the regime $V > -2\epsilon_0$, where effect (I) is canceled and no pairing is expected from mean-field theory. In contrast to the common wisdom, a collaboration between the kinetic energy and the restricted Hilbert space from effect (II) itself can induce a strong pairing. 
Because the $\Delta_d$ splitting penalizes spin-triplet doublon states, single-electron motion is frustrated. Remarkably, it turns out in the large $\Delta_d$ limit, forming a Cooper pair can lower the kinetic energy within the constrained low-energy subspace.  In this mechanism the $T_c$ will be decided by the intra-layer hopping.

\begin{figure}
    \centering
\includegraphics[width=1\linewidth]{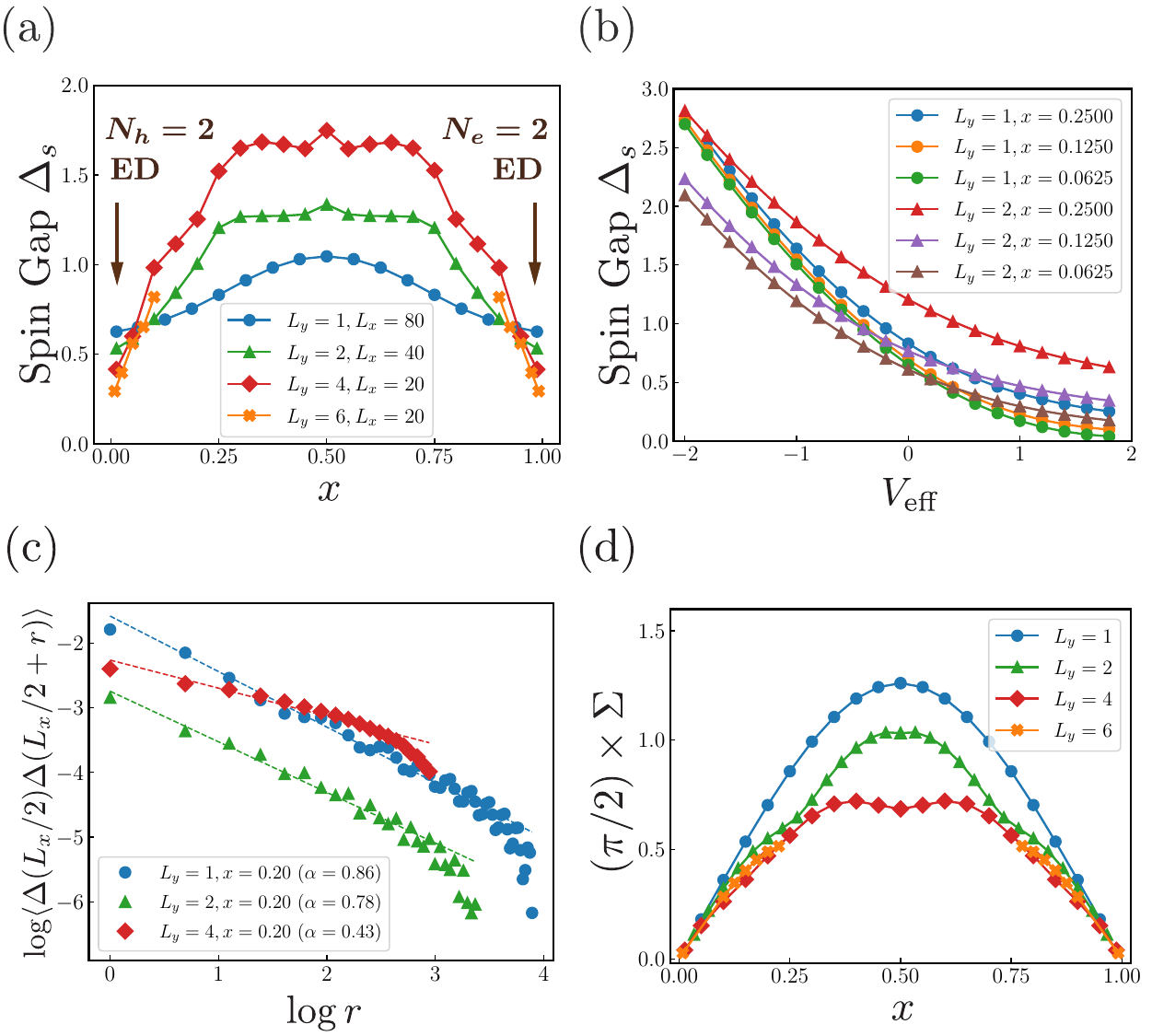}
    \caption{\textbf{Evidence of kinetic superconductivity from DMRG simulation of ESD model.}
    (a) Doping dependence of the spin gap at $r = 1$ and $\textcolor{black}{V_{\mathrm{eff}}} = 0$ for different system sizes ($L_y = 1, 2, 4, 6$) with bond dimension $\chi = 3000$. A dome-like behavior of the spin gap as a function of $x$ is consistently observed, although the quantitative values vary with system size $L_y$. 
(c) Dependence of the spin gap on $\textcolor{black}{V_{\mathrm{eff}}}$ at $r = 1$. Different markers denote different system sizes: $L_y = 1$, $L_x = 80$ (circles) and $L_y = 2$, $L_x = 40$ (triangles).
(c) Pair-pair correlation function at $x = 0.2$ for various $L_y = 1, 2, 4$ with $\chi = 3000$, using $r = 1$ and $\textcolor{black}{V_{\mathrm{eff}}} = 0$. All pair correlation functions exhibit power-law decay with an exponent $\alpha$ that decreases with increasing $L_y$.
(d) Doping dependence of the superfluid weight $D_s$ at $r = 1$ and $\textcolor{black}{V_{\mathrm{eff}}} = 0$. We use $L_x = 40$ ($L_y = 1$), $L_x = 30$ ($L_y = 2$), and $L_x = 20$ ($L_y = 4, 6$) with $\chi = 3000$. 
 Reprinted from Ref.~\cite{oh2025hightemperaturesuperconductivitykineticenergy}.   
    }
    \label{fig:esd_dmrg}
\end{figure}

\begin{figure}
    \centering
\includegraphics[width=1\linewidth]{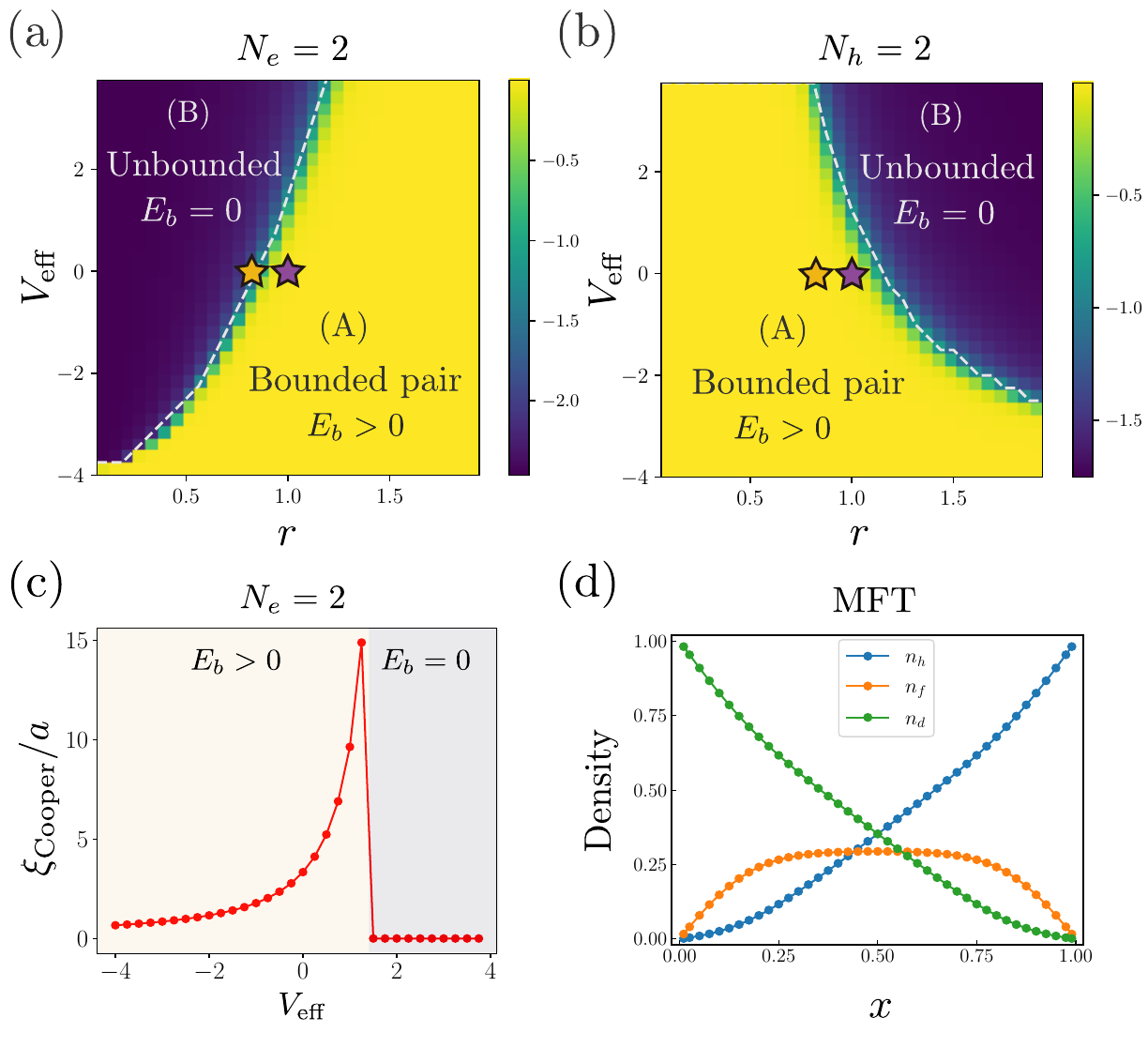}
    \caption{\textbf{Evidence of kinetic superconductivity from exact solution and mean-field theory of ESD model.} (a,b) Phase diagrams in the $(r,\textcolor{black}{V_{\mathrm{eff}}})$ for the $N_e = 2$ and $N_h = 2$ cases. The color bar represents the system-size scaling exponent $\alpha$, where $E_b \sim L^{-\alpha}$. For $\textcolor{black}{V_{\mathrm{eff}}} < \textcolor{black}{V_{\mathrm{eff},c}}$ (yellow region), the bound-state energy is finite, while for $\textcolor{black}{V_{\mathrm{eff}}} > \textcolor{black}{V_{\mathrm{eff},c}}$ (black region), the bound-state energy converges to zero. We denote the points $r = 1$, $\textcolor{black}{V_{\mathrm{eff}}} = 0$ and $r = \sqrt{2}/\sqrt{3}$, $\textcolor{black}{V_{\mathrm{eff}}} = 0$ with purple and yellow stars, respectively. 
(c) The size of Cooper pair as a function of $V_{\mathrm{eff}}$ at $r=1$. The cooper pair size is evaluated from $\xi_{\mathrm{Cooper}}=\sqrt{R^2|\Psi^{(2e)}(R)|^2}$ in the exact diagonalzation method with system size $L_{x}=L_y=100$. (d) Mean-field solutions of the ESD model in the two-dimensional limit. Here, we use $L_x=L_y=40$. Reprinted from Ref.~\cite{oh2025hightemperaturesuperconductivitykineticenergy}.
}
    \label{fig:esd_ed}
\end{figure}

The presence of superconductivity was demonstrated by three  independent methods~\cite{oh2025hightemperaturesuperconductivitykineticenergy}. Below, we summarize the key findings from each approach:
\begin{itemize}
    \item \textbf{DMRG simulation results}: Density Matrix Renormalization Group (DMRG) simulations provide strong evidence for a Luther-Emery liquid ground state across a broad doping range $x$ in the ESD model at $\textcolor{black}{V_{\mathrm{eff}}} = 0$. As illustrated in Fig.~\ref{fig:esd_dmrg}, this evidence includes a finite spin gap and charge-$1e$ gap, a vanishing charge-$2e$ gap, power-law pair–pair correlations, and a finite phase stiffness $D_s$. Interestingly, superconductivity remains very robust even under repulsive interactions $V_{\mathrm{eff}}>0$, as shown in Fig.~\ref{fig:esd_dmrg}(b). For $L_y$ from $1$ to $6$, there is a dome for both the pairing gap and the phase stiffness. Both the pairing gap and superfluid weight can reach order one of the intra-layer hopping $t$, suggesting a really strong superconductor with $T_c \sim t$ at the optimal doping.

    \item \textbf{Exact solution of two particle problem}: In the dilute limit ($N_e = 2$ or $N_h = 2$), the ESD model can be mapped onto a single-particle hopping problem with a non-uniform impurity located at the origin, which can be exactly solved in the 2D limit. The exact solution reveals a finite binding energy $E_b>0$, even in the absence of explicit attractive interactions (See Fig.~\ref{fig:esd_ed}~(a-b)).  
    This indicates the formation of bound Cooper pairs purely from kinetic processes.
This analysis highlights the crucial role of the unconventional hopping term (see Fig.~\ref{fig:esd_pd}~(d)), which introduces bond-disorder-like terms that enable bound state formation. Furthermore, the size of the Cooper pair, $\xi_{\mathrm{Cooper}} $, is found to be large (See Fig.~\ref{fig:esd_ed}~(c)), indicating a spatially extended, weakly bound pair and clearly distinct from tightly bound BEC-like pairs.

\item \textbf{Slave-boson mean field results}: A generalized slave-boson mean-field theory~\cite{oh2025hightemperaturesuperconductivitykineticenergy} qualitatively reproduces the phase diagram and confirms the pairing mechanism arising purely from the kinetic energy (See Fig.~\ref{fig:esd_ed}~(d)). The results highlight the importance of the unconventional hopping terms, unique to the ESD model, that drive kinetic pairing and also make connection to the Feshbach resonance widely discussed in the cold atom systems.  Similar Feshbach resonance pictures were also discussed in Ref~\cite{lange2023pairing,Schlömer2024,PhysRevB.109.045127}. 
\end{itemize}

\subsection{Normal state as second Fermi liquid}
In hole doped cuprates, one mystery is the exotic pseudogap normal state at small doping $x$.  In the ESD model, Ref.~\cite{yang2024strong,oh2025hightemperaturesuperconductivitykineticenergy} also propose a similar but simpler ``pseudogap metal'' at small $x$, which is disconnected to the conventional Fermi liquid at large $x>0.5$.  In the ESD model, there are two different vacuums in the $x=0$ and $x=1$ limit, which  correspond to a product of $\ket{d}$ states and a product of $\ket{h}$ states, respectively. The normal states near $x=0$ and $x=1$ can be understood as doping these two different vacuums and  realize two distinct Fermi liquids (FLs). Specifically,  the Fermi-surface volume per flavor is given by $A_{FS} = -x/2$ at small doping $x\rightarrow 1$, while at large doping $x\rightarrow 1$ it is $A_{FS} = (1-x)/2$. 
The first case violates the conventional Luttinger theorem for the Fermi volume by half of the Brillouin zone per flavor~\cite{PhysRevLett.84.3370,yang2024strong}, leading to what is dubbed as the second Fermi liquid (sFL).  The sFL phase violates the perturbative form of the Luttinger theorem but remains consistent with Oshikawa’s non-perturbative proof~\cite{oshikawa2000topological}. We emphasize that this metallic state is an intrinsically strongly correlated Fermi liquid that lies beyond the any weak-coupling theory and is closely connected to the concept of symmetric mass generation~\cite{sym14071475}. The nature of the transition between sFL and FL is an open question. It involves a jump of the Fermi surface volume by half of the BZ and is definitely different from a Lifshitz transition.  Strange metal with linear $T$ behavior has been conjectured to arise in similar Fermi surface volume change transition in heavy Fermion systems and in hole doped cuprate~\cite{RevModPhys.75.913,RevModPhys.92.011002}.  It is interesting to study whether a strange metal phase arises in the ESD model around $x=0.5$.

\section{Phenomenology of bilayer nickelate} \label{Sec5}

We now move to specific realization in realistic materials. The bilayer Kondo model and type II t-J model discussed above may be relevant to the recently found bilayer nickelate La$_3$Ni$_2$O$_7$ system~\cite{sun2023signatures}. In this section, we briefly summarize the existing experimental results and theoretical discussions of this new material. We note that superconductivity was also observed in trilayer nickelate, La$_4$Ni$_3$O$_{10}$~\cite{Zhu2024,zhang2025superconductivity,Li_2024_tri}, but we will restrict to the bilayer material.  
\begin{figure}
    \centering
\includegraphics[width=1\linewidth]{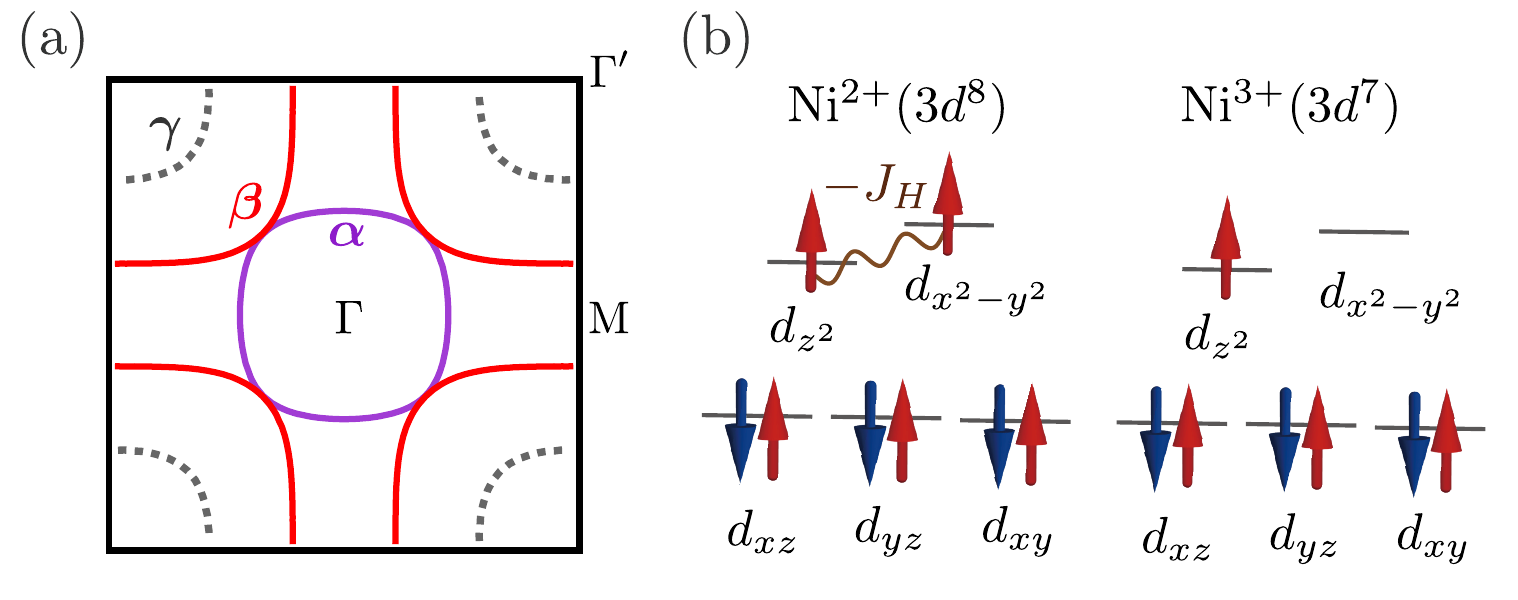}
    \caption{\textbf{Illustration of schematic Fermi surface and electron filling in Ni$^{2+}$ and Ni$^{3+}$}. 
(a) The $\alpha$, $\beta$, and $\gamma$ Fermi pockets predicted by DFT calculation under high pressure~\cite{luo2023bilayer} are depicted. Some of ambient pressure ARPES experiments detect only the $\alpha$ and $\beta$ pockets dominated by the $d_{x^2-y^2}$ orbital, while the $\gamma$ pocket is absent, suggesting that the band from the $d_{z^2}$ orbital lies below the Fermi level~\cite{zhou2024revealing,Yang2024,wang2025electronic,shen2025anomalous}. (b) The average Ni valence in bilayer nickelates is Ni$^{ (2+x)+}$, with $x=0.5$ corresponding to a 3$d^{7.5}$ configuration. 
Reprinted from Ref.~\cite{oh2025highspinlowspin}. 
}
    \label{fig_fermi_surface}
\end{figure}

\subsection{Summary of experimental results}

Superconductivity with $T_c \approx 80$ K was reported in the high pressure sample of bulk La$_3$Ni$_2$O$_7$ systems~\cite{sun2023signatures}. In addition to vanishing resistivity~\cite{sun2023signatures,Zhang2024_str,Hou_2023}, diamagnetism measurement also indicated signatures of the Meissner effect~\cite{10.1093/nsr/nwaf220,Wang2024}.  More recently, superconductors with $T_c \approx 40 $ K were also realized at ambient pressure in thin films with strain induced by substrate~\cite{Ko2025,Zhou2025,Huo2025}, which opens the window to probe the system with more accessible measurements, including ARPES and STM~\cite{Li_2024,wang2025electronic,shen2025anomalous,sun2025observation,fan2025superconducting}. The field is still rapidly developing, but a lot of information has already been revealed, which we summarize below:

\begin{itemize}
    \item  The superconducting phase is mainly from a  block of bilayer square lattice.  Other structures such as monolayer-trilayer configuration have also been discovered~\cite{PhysRevLett.133.146002}, but it remains to see whether superconductivity exists in such stacking.

\item The essential low-energy physics involves the $d_{x^2-y^2}$ and $d_{z^2}$ orbitals of Ni.  Angle-resolved photoemission spectroscopy (ARPES) measurements have been performed on both unpressurized bulk crystals and strained thin films.  
In bulk samples, Ref.~\cite{Li_2024,Yang2024} reported two Fermi surfaces ($\alpha$ and $\beta$ pockets) of $d_{x^2-y^2}$ character, while the $\gamma$ pocket predicted by DFT calculation~\cite{luo2023bilayer} was absent and the associated $d_{z^2}$ band lies below the Fermi level(see Fig.~\ref{fig_fermi_surface}(a)).   
For thin films, the situation remains unsettled. Some ARPES studies report the $\gamma$ pocket~\cite{10.1093/nsr/nwaf205,shen2025anomalous}, whereas others~\cite{wang2025electronic,sun2025observation} did not see the $\gamma$ pocket.  
Clarifying this issue—and establishing whether the $\gamma$ pocket plays an essential role in superconductivity—remains an important open question for future experiments.

    \item The average valence of Ni in the current experiment is at Ni$^{(2+x)+}$ with $x=0.5$ at each site. One natural expectation is that Ni is in the $3d^{8-x}$ configuration with $x=0.5$ (see Fig.~\ref{fig_fermi_surface}(b)). However, it is possible that the system is in the charge transfer regime instead of the Mott-Hubbard regime.   There exists experimental indication that the system is in the charge transfer regime~\cite{Dong2024}. In this scenario, starting from a parent state at $3d^{8}$ valence, the doped holes enter the oxygen $p$ orbitals instead of Ni $3d$ orbitals. We note that this issue may not matter too much for low energy descriptions. From the lesson in the hole doped cuprates, the low energy effective models are often the same in the Mott-Hubbard regime and charge transfer regime.
    
    \item Recent ARPES measurements have also made progress in resolving the pairing gap directly in the superconducting state indicating the absence of a gap node~\cite{sun2025observation,shen2025anomalous}. Consistently, a STM measurement on thin film sample revealed a two-gap structure in the tunneling spectrum, suggesting that the superconducting pairing is fully gapped without a node~\cite{fan2025superconducting}. These results are consistent with  simple s-wave pairing, but phase sensitive measurement is needed to directly verify the angular momentum of the pairing.

    \item  In unpressured bulk samples and compressively or tensile strained thin films, evidences for spin-density-wave (SDW) order with momentum $Q=(\tfrac{\pi}{2},\tfrac{\pi}{2})$ have been reported in resonant inelastic x-ray scattering (RIXS) measurements~\cite{Chen2024,gupta2024anisotropicspinstripedomains,Ren2025}. 
Consistent signatures of SDW order have also been observed using complementary probes, including muon-spin rotation/relaxation experiments~\cite{PhysRevLett.132.256503,Khasanov2025}, nuclear magnetic resonance (NMR)~\cite{FUKAMACHI2001195,doi:10.7566/JPSJ.93.053702,Zhao_2025}, and neutron scattering measurements~\cite{plokhikh2025unravelingspindensitywave}.

Several experiments found that the SDW order is suppressed under pressure or strain before the onset of the superconductivity.  But it remains to see whether the superconductivity can coexist with the SDW order. Interestingly, one recent transport experiment suggests a coexistence of superconductivity and a spin glass order~\cite{ji2025signaturesspinglasssuperconductivitynickelate}.  
    
\item There have been several transport measurements of the Hall coefficient on thin-film bilayer nickelates.  
The Hall coefficient is highly sensitive to oxygen stoichiometry and having some variations among different group.
Refs.~\cite{Liu2025,Zhou2025,wang2025electron} reported that the Hall coefficient is significantly smaller than that in cuprate superconductors, whereas Refs.~\cite{Ko2025,Liu2025} found values comparable to those of overdoped cuprates.  
Such contrasting observations are difficult to reconcile within a simple single-band framework pointing out the effects of multiband electronic structure with complicated carrier dynamics.

\item The normal state of the superconductor in the pressurized bulk sample has a linear $T$ resistivity~\cite{sun2023signatures,Zhang2024_str,Hou_2023}, similar to the strange metal at the optimal doping of the hole doped cuprates. In contrary, the normal state of the superconductor in the strained thin films is a Fermi liquid with $T^2$ resistivity~\cite{Liu2025,zhou2024revealing,Hao2025}. 

\item  It is commonly reported that the superconducting transition in bilayer nickelates is accompanied by a structural phase transition~\cite{sun2023signatures,Zhang2024_str,Hou_2023}. While the normal state at low pressure is generally assigned to the orthorhombic \textit{Amam} phase, there remains an ongoing debate as to whether the superconducting state corresponds to the orthorhombic \textit{Fmmm} structure or whether an additional transition occurs into the tetragonal \textit{I4/mmm} phase~\cite{10.1093/nsr/nwaf220,Wang2024_str}. Furthermore, whether and how the underlying crystal symmetry influences the emergence of superconductivity remains unclear.

\item The doping dependence remains comparatively less explored. Most of the studies focus on a fixed hole doping $x=0.5$ per site relative to the $3d^8$ valence.  A recent experiment explored  Sr$^{2+}$ doped thin films with $x=0.5+\delta$. It found that $T_c$ persists  over a wide doping range ($0 < \delta < 0.21$)~\cite{Hao2025}. It is highly encouraging to electron dope the sample, such as substituting Ce, to achieve $x<0.5$  and  establish a complete doping-dependent phase diagram.

\item Recent efforts have attempted to bridge up between high-pressure studies and thin-film studies in bilayer nickelates.  In particular, people try to put thin films under pressure and observed enhancement of $T_c$ ~\cite{Osada2025,li2025enhanced}.  Ref.~\cite{Osada2025} also reported 
some results about $T_c$ dependence on pressure and lattice constant. 

\end{itemize}

\subsection{Summary of existing theoretical studies}
In parallel with these experimental discoveries, a wide range of theoretical studies has been done including the functional renormalization group~\cite{yang2023possible,cao2025strain,zhan2024cooperation,gu2023effective,cao2025strain,le2025landscape}, random phase approximation~\cite{liu2023s}, fluctuation exchange~\cite{sakakibara2023possible,PhysRevB.111.104505,ushio2025theoretical,PhysRevB.109.104508}, dynamical mean-field theory~\cite{tian2024correlation,chen2024non}, auxiliary-field Monte Carlo~\cite{qin2023high}, and the slave-boson approach in the strong-coupling limit~\cite{oh2023type,lu2023interlayer,ji2025strong}. Most of these works mainly focus on the magnetic interactions as a source of the pairing glue, but some of them also include the electron–phonon coupling~\cite{zhan2024cooperation,yin2025s}. In addition, a pairing dome arising from a Feshbach resonance in the strong-coupling regime has also been proposed~\cite{lange2023pairing,yang2024strong,oh2025hightemperaturesuperconductivitykineticenergy,PhysRevB.109.045127}.
In terms of pairing symmetry and orbital content of the charge carriers, there have been various proposals listed as the followings:
\begin{enumerate}
    \item \textbf{Interlayer $s$-wave pairing of $d_{x^2-y^2}$ orbital:}  Some works assume that the $d_{z^2}$ orbital is Mott localized.  But the strong inter-layer super exchange coupling $J_\perp$ of $d_{z^2}$ orbital can be shared to the $d_{x^2-y^2}$ orbital through the Hund's coupling $J_H$. In the end, this effective $J_{\perp;\mathrm{eff}}$ coupling drives an 
 interlayer $s$-wave pairing for the $d_{x^2-y^2}$ orbital~\cite{lu2023interlayer,oh2023type,qu2023bilayer,lange2023pairing,PhysRevB.109.045127,lu2023superconductivitydopingsymmetricmass,duan2025orbital,zhang2023strong}.

   \item \textbf{Intra-layer $d$-wave pairing of $d_{x^2-y^2}$ orbital:}  In the scenario of $d_{z^2}$ to be Mott localized, the mobile carriers are in the $d_{x^2-y^2}$ orbital. In principle, there is a possibility of intra-layer $d_{x^2-y^2}$ pairing just as in cuprates~\cite{zhan2024cooperation}.  One may expect a transition from the intra-layer d-wave pairing to inter-layer s-wave pairing by increasing the ratio $J_\perp/J$, where $J$ is the intra-layer super-exchange of the $d_{x^2-y^2}$ orbital~\cite{oh2023type,lu2023interlayer,duan2025orbital,PhysRevB.110.094509,tian2025spin}. 
   We note that intra-layer d-wave pairing implies a  pairing node which has not been found in the existing measurements~\cite{shen2025anomalous,fan2025superconducting}.

   \item \textbf{$s$-wave pairing of $d_{z^2}$ orbital:}  Some works assume that the mobile carrier is mainly in the $d_{z^2}$ orbital~\cite{PhysRevB.108.214522,zhu2025quantum,sakakibara2023possible,shen2023effective,yang2023minimal}. One can imagine a parent state with $d_{z^2}$ orbital in a rung-singlet phase, then hole doping simply mobilizes the pairing and results in phase coherence. 

   \item \textbf{ Mixture of $d_{z^2}$ and $d_{x^2-y^2}$ orbitals:} In one scenario, charge carriers enter both the $d_{z^2}$ and $d_{x^2-y^2}$ orbitals. In this case, the inter-site hybridization between the two orbitals can transfer the inter-layer pairing from the $d_{z^2}$ orbital to the $d_{x^2-y^2}$ orbital~\cite{yang2023minimal}.

    \item \textbf{$s_{\pm}$-wave pairing in weak coupling:} 
    Weak-coupling approaches such as RPA and fRG predict an $s_{\pm}$-wave state, characterized by a sign change of the superconducting gap between the $\alpha$ and $\beta$ Fermi-surface pockets of bilayer nickelates ~\cite{Zhang2024,zhang2023trends,PhysRevB.109.104508,PhysRevB.108.L140505,PhysRevLett.131.236002,PhysRevB.110.195135}.  We note the sign change simply indicates that inter-layer pairing dominates over the intra-layer pairing. Therefore, the  pairing symmetry in this scenario can smoothly crossover to the inter-layer s-wave pairing in the strong coupling limit. 
  
    \end{enumerate}
Beyond superconductivity, numerous theoretical studies have also been devoted to SDW order or magnetic correlation~\cite{liu2025origin,PhysRevB.110.205122,PhysRevB.112.024508}. 
Overall, theoretical studies of this system remain highly active and continue to evolve.

\subsection{A possible theory and its limitations}
Here we offer one theory of the bilayer nickelate superconductivity based on the picture of doped spin-one Mott insulator developed in Sec. \ref{Sec3} and Sec. \ref{Sec4}. In this framework, the $d_{z^2}$ orbital is assumed to be Mott localized and only provides a spin moment.  This is consistent with the ARPES results indicating the absence of the $\gamma$ pocket from the $d_{z^2}$ orbital.  Then the low energy effective model is the bilayer FM Kondo model or the bilayer type II t-J model when $J_H$ is large. Then from the results in Sec.~\ref{Sec3} and Sec.~\ref{Sec4}, we know there exists a robust superconductor with inter-layer s-wave pairing symmetry in the large $J_\perp$ limit.  The pairing survives even if there is an inter-layer repulsion $V$ which cancels the net attractive interaction from $J_\perp$.  We note s-wave pairing is consistent with the ARPES\cite{sun2025observation,shen2025anomalous} and STM measurements\cite{fan2025superconducting} indicating a full gap. In large $J_\perp$ limit, we expect a phase diagram illustrated in Fig.~\ref{fig:esd_pd}(c).  We expect that the pairing gap has a dome dependence on $x$, with the optimal doping $x_c$ close to $0.5$. This may offer a plausible explanation for why a relatively high $T_c$ is observed experimentally at such a large hole doping $x=0.5$. At the same doping level in cuprate, superconductivity is known to be absent. Therefore the doping dependence of $T_c$ can reveal  new physics distinct to the familiar cuprate family.  Based on this theory, it is interesting to further electron dope the system to reach the true optimal doping and study the possible non-Fermi-liquid illustrated in Fig.~\ref{fig:esd_pd}(c).

Although the model and theory presented in Sec.~\ref{Sec3} and Sec.~\ref{Sec4} account for essential experimental observations, they possess several limitations that must be addressed for a more comprehensive description. These limitations are as follows:

\begin{enumerate}
    \item  In the experiment, the unpressured or unstrained sample exhibits a SDW order with $Q=(\frac{\pi}{2},\frac{\pi}{2})$. Such a magnetic order is absent in the analysis in the large $J_\perp$ regime provided in Sec.~\ref{Sec4}. We conjecture that the magnetic order emerges in the $J_\perp \rightarrow 0$ limit, where we can focus on each layer described by a FM Kondo model or type II t-J model.  It is known that the FM Kondo model hosts a FM phase with $Q=(0,0)$ in the large  $J_H$ regime due to the double exchange mechanism~\cite{PhysRev.118.141}.  In our model there is also in-plane super-exchange $J>0$ which favors Neel order with $Q=(\pi,\pi)$. The competition between the ferromagnetic double exchange and antiferromagnetic $J$ may result in a magnetic order with a large period. 

   We conjecture that the SDW order in the experiment can be accessed in future numerical study of either the FM Kondo model or type II t-J model with some variations. In this picture, the local moment from $d_{z^2}$ orbital is crucial, so the magnetic phase likely cannot be explained in  models including only the $d_{x^2-y^2}$ orbital. In the future it is interesting to study the evolution under increasing $J_\perp/J$ in the bilayer FM Kondo model or type II t-J model.  Right now it is not clear whether there is a direct transition from the SDW phase to the superconductor or an intermediate coexisting phase.

\item In the models studied in Sec.~\ref{Sec3} and Sec.~\ref{Sec4}, there is no inter-layer hopping $t_\perp$ for the $d_{x^2-y^2}$ orbital.  While the direct inter-layer hybridization for $d_{x^2-y^2}$ is indeed small, there could be an effective $t_\perp$ generated from virtual process to the gapped $d_{z^2}$ orbital.  We expect such a term has the form $t_{\perp;\mathrm{eff}} (\cos k_x -\cos k_y)^2$ in momentum space and generate the $\alpha,\beta$ pockets for the $d_{x^2-y^2}$ orbital. The term can be easily included in the models. It remains to study the effect of such a  term. The hybridization of the two Fermi surfaces is likely important to explain the small Hall coefficient because $\alpha, \beta$ pockets are electron-like and hole-like respectively with a sizable $t_{\perp;\mathrm{eff}}$. However, we conjecture that it may not influence the superconductivity too much.

\item  In the current models, the $d_{z^2}$ orbital is assumed to be Mott localized, fixing the electron densities at each site to $n_1=1-x$ for the $d_{x^2-y^2}$ orbital and $n_2=1$ for the $d_{z^2}$ orbital. However, depending on system parameters and doping, the $d_{z^2}$ orbital can also become slightly hole-doped. This scenario modifies the densities to $n_1=1-(x-\delta)$ and $n_2=1-\delta$, where $\delta$ is the small hole density in the $d_{z^2}$ orbital. This regime can be studied using an extended type-II t-J model with seven states per site~\cite{oh2023type}. While a small $\delta$ is unlikely to cause qualitative changes to the phase diagram, this should be confirmed explicitly via numerical simulation. 

Another possibility is that the system lies in the charge-transfer regime of the Zaanen--Sawatzky--Allen classification~\cite{PhysRevLett.55.418}, where the charge-transfer energy is much smaller than the Hubbard $U$. In this case, the ligand oxygen $2p$ orbitals should participate in the low-energy physics together with the nickel $d$ orbitals. Ref.~\cite{PhysRevB.111.L020504} shows that the minimal model in this regime remains a bilayer type-II $t$--$J$ model, incorporating the Zhang--Rice doublet state ($d^8L$). However, a comprehensive study of the interplay between pressure or the role of the outer-plane oxygen is still needed.

\item Currently, robust DMRG evidence for superconductivity in wider systems ($L_y$ up to $8$) has been obtained for the ESD model, which is applicable only in the large $J_\perp$ regime~\cite{oh2025hightemperaturesuperconductivitykineticenergy}. Although the pairing is expected to persist for finite $J_\perp/t$, this has only been explicitly confirmed for small $L_y$~\cite{oh2025hightemperaturesuperconductivitykineticenergy}. Future numerical work should therefore focus on the bilayer FM Kondo or type-II t-J models with large $L_y$ to systematically determine the dependence of the pairing strength on $J_\perp$.

\item Analytically, the primary method for describing the pairing involves a mean-field decoupling of the interlayer $J_\perp$ term~\cite{oh2023type,lu2023interlayer}. This approach suggests a mechanism driven by an effective attractive interaction. However, this picture is challenged by numerical results showing that pairing exists even when the net interaction is repulsive (see Fig.~\ref{fig:pd_bilayer_typeII}(c)). This discrepancy indicates that a simple mean-field theory may be  insufficient to capture the essential physics.   Also, at small $x$, the normal state likely transits from the conventional Fermi liquid (FL) to the second Fermi liquid (sFL) when increasing $J_\perp$, accompanying a jump of Fermi surface volume~\cite{wu2024deconfined}. 
Therefore a successful theory should capture the pairing mechanism for both normal states through the  transition. Developing a more comprehensive theoretical framework that works for general values of $J_\perp$ and $V$ remains a critical challenge.
\end{enumerate}

\section{Conclusion}
In this review, we have summarized recent theoretical progress in understanding doped spin-one Mott insulators with a $3d^8$ electronic configuration. The central theoretical framework is an effective FM Kondo or type-II t-J model, which arises when the $d_{z^2}$ orbitals are Mott localized and the $d_{x^2-y^2}$ orbitals contain mobile carriers. While the FM Kondo lattice model typically hosts an FM phase via the double-exchange mechanism, we highlighted that superconductivity can emerge when the parent insulator is in a  paramagnetic phase, such as a spin-one Haldane chain or a two-leg ladder. Our primary focus was the models on a bilayer square lattice, which may be of direct relevance to the recently discovered bilayer nickelate superconductor La$_3$Ni$_2$O$_7$. In this system, a large inter-layer spin coupling ($J_\perp$) for the $d_{z^2}$ orbitals suppresses magnetism, instead favoring a robust, inter-layer s-wave superconductor for the mobile electrons in the $d_{x^2-y^2}$ orbital. The models and theories discussed provide a compelling lens through which to view the existing experimental results  and to chart a course for future exploration.

Beyond explaining current observations, the bilayer models present tantalizing possibilities that could offer new insights into the enduring puzzle of high-$T_c$ superconductivity. We highlight two particularly promising future directions:

\begin{enumerate}
    \item \textbf{Engineering superconductor with higher Tc:} The bilayer model predicts that in the limit of large $J_\perp$, the superconducting transition temperature ($T_c$) can approach the order of the intra-layer hopping energy ($t$). This pairing is remarkably resilient, scaling with hopping and persisting even against significant repulsive interactions. While the current nickelate La$_3$Ni$_2$O$_7$ may not be in this ideal regime, it points to a clear materials-design principle: enhancing $J_\perp$, perhaps by reducing the c-axis lattice constant, could be a pathway to achieving exceptionally high-temperature superconductivity.

    \item \textbf{A Clean Platform for Non-Fermi-Liquid Physics:} In the strong $J_\perp$ regime of the model, the underdoped normal state manifests as a symmetric pseudogap (PG) metal with small hole pockets. This phase is a strongly interacting Fermi liquid, distinct from the conventional Fermi liquid  at higher doping. The transition between these two metallic states offers a pristine theoretical arena to study non-Fermi-liquid criticality. Unhindered by the complexities of symmetry breaking that often cloud the physics in materials like the cuprates and heavy fermion materials, the bilayer model provides an ideal setting to explore one of the most challenging problems in condensed matter physics.  The criticality may also be experimentally realized in the bilayer nickelate through electron doping or increasing the pressure.
\end{enumerate}

\textbf{Acknowledgement}: We thank Annabelle Bohrdt, Debanjan Chowdhury, Fabian Grusdt, Srinivas Raghu, Kyle Shen, Qimiao Si, Ashvin Vishwanath, Yijun Yu, Jing-Yu Zhao and Boran Zhou for discussions related to bilayer nickelate. The work is supported by a startup fund from the Johns Hopkins University.

%

\onecolumngrid
\newpage
\clearpage
\setcounter{equation}{0}
\setcounter{table}{0}
\setcounter{page}{1}
\setcounter{section}{0}

\maketitle 
\makeatletter
\renewcommand{\theequation}{S\arabic{equation}}
\renewcommand{\thefigure}{S\arabic{figure}}
\renewcommand{\thetable}{S\arabic{table}}
\renewcommand{\thesection}{S\arabic{section}}

\appendix
\onecolumngrid

\end{document}